\documentclass[10pt,aps,prx,footinbib,floatfix,superscriptaddress,
	       twocolumn,showpacs,longbibliography]{revtex4-1} 

\usepackage{graphicx}
\usepackage{dcolumn}
\usepackage{bm}
\usepackage[utf8]{inputenc}
\usepackage{amssymb}
\usepackage{amsmath}
\usepackage{pstricks}
\usepackage{color}
\usepackage{slashed}
\usepackage{braket}
\usepackage{textcomp}
\usepackage{hyperref}
\usepackage{natbib}
\usepackage{extarrows}
\usepackage{wasysym}
\usepackage[normalem]{ulem}
\usepackage{verbatim}
\usepackage{physics}

\newcommand{\cohfact}{\langle \cos(\varphi) \rangle}
\newcommand{\um}{\kern 0.2em{\textmu}m}

\newcommand{\reddiamond}{\protect\includegraphics[height= 8 pt]{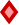}}
\newcommand{\bluecircle}{\protect\includegraphics[height= 6 pt]{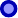}}
\newcommand{\purpletriangle}{\protect\includegraphics[height= 6 pt]{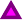}}
\newcommand{\orangesquare}{\protect\includegraphics[height= 6 pt]{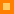}}

\newcommand{\redsquare}{\protect\includegraphics[height= 6 pt]{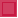}}
\newcommand{\bluediamond}{\protect\includegraphics[height= 8 pt]{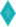}}
\newcommand{\orangecircle}{\protect\includegraphics[height= 6 pt]{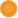}}

\definecolor{azure}{rgb}{0.0, 0.5, 1.0}
\definecolor{amber}{rgb}{1.0, 0.49, 0.0}
\definecolor{forestgr}{rgb}{0.13, 0.55, 0.13}

\begin{document}
\title{Extracting the field theory description of a quantum many-body system from experimental data}
\author{Torsten V.~Zache}
\affiliation{Institut f\"ur Theoretische Physik, Ruprecht-Karls-Universit\"at Heidelberg, Philosophenweg~16, 69120~Heidelberg, Germany}
\author{Thomas Schweigler}
\affiliation{Vienna Center for Quantum Science and Technology, Atominstitut, TU Wien, Stadionallee 2, 1020 Vienna, Austria}
\author{Sebastian Erne}
\affiliation{Vienna Center for Quantum Science and Technology, Atominstitut, TU Wien, Stadionallee 2, 1020 Vienna, Austria}
\affiliation{School of Mathematical Sciences, Centre for the Mathematics and Theoretical Physics of Quantum Non-Equilibrium Systems, University of Nottingham, University Park, Nottingham NG7 2RD, UK}
\author{J{\"o}rg Schmiedmayer}
\affiliation{Vienna Center for Quantum Science and Technology, Atominstitut, TU Wien, Stadionallee 2, 1020 Vienna, Austria}
\author{J{\"u}rgen Berges}
\affiliation{Institut f\"ur Theoretische Physik, Ruprecht-Karls-Universit\"at Heidelberg, Philosophenweg~16, 69120~Heidelberg, Germany}

\date{\today}

\begin{abstract}
Quantum field theory is a powerful tool to describe the relevant physics governing complex quantum many-body systems. Here we develop a general pathway to extract the irreducible building blocks of quantum field theoretical descriptions and its parameters purely from experimental data. This is accomplished by extracting the one-particle irreducible (1PI) correlation functions from which one can construct all physical observables. To match the capabilities of experimental techniques, our approach employs a formulation of quantum field theory based on equal-time correlation functions only. We illustrate the theoretical foundations of our procedure by applying it to the sine-Gordon model in thermal equilibrium, and then demonstrate explicitly how to extract these quantities from an experiment where we quantum simulate the sine-Gordon model by two tunnel-coupled superfluids. We extract all 1PI correlation functions up to the 1PI four-point function (interaction vertex) and their variation with momentum, encoding the `running' of the couplings. The measured 1PI correlation functions are compared to the theoretical estimates, verifying our procedure. 
Our work opens new ways of addressing complex many-body questions emerging in a large variety of settings from fundamental science to practical quantum technology.
\end{abstract}

\maketitle

Quantum Field Theory (QFT) has a wide range of very successful applications from early-universe cosmology and high-energy physics to condensed matter physics. A central aspect of QFT is that it describes the many-body limit of complex interacting quantum systems, which is also relevant for quantum technology if devices become large. 
Present large-scale analog quantum simulators using ultra-cold atoms explore the many-body limit described by QFT, e.g.~\cite{Bloch2008,haller2010,Gring12,Hung2013,Langen15,Navon2015,Navon2016,parsons2016site,schweigler2017experimental,bernien2017probing,Pruefer2018,Erne2018,Eckel2018,Hu2019,Feng2019,Murthy2019,Keesling2019,prufer2019experimental}. 
Therefore, they may also be used to solve outstanding theoretical problems of QFT that are beyond classical computational techniques.
 
One of the big experimental challenges is probing the complex many-body states. One strategy is to detect every constituent (atom, superconducting qubit, quantum dot \dots) and its state. Such detections constitute a projective measurement of the many-body wave function in the constituent basis. For large systems such a measurement contains way too much information to be ever analysed fully. This is reflected by the exponential complexity of ‘tomography’ that prevents a complete characterization of the many-body quantum states~\cite{flammia2012quantum}.
 
By contrast, there are important simplifications occurring in the many-body limit described by QFT. In QFT, often only a small subset of the microscopic details of the underlying theory is relevant for the computation of measurable physical properties. This effective loss of details has its mathematical foundation in the renormalization program of QFT~\cite{weinberg1995quantum}. As a result, for a quantum simulation of such a theory, many of the detailed properties of the microscopic quantum device have no effect on the simulation outcome for quantities of interest~\cite{berges2019scalingup}.

This raises the important question of how to extract from experimental data the relevant information content of QFT. It is well known that an efficient description of QFT can be based on one-particle irreducible (1PI) correlation functions, called irreducible or proper vertices~\cite{weinberg1995quantum}. They represent the irreducible building blocks from which all physical observables may be constructed. This can be, e.g. the effective Hamiltonian determining the macroscopic dynamics, a possible spectrum of quasi-particles and their effective interaction strength. In a general setting, these vertices are functions of space and time or momentum and frequency, encoding the ‘running' of couplings prominently discussed in high-energy physics in the framework of the Standard Model of particle physics.
 
In principle, the irreducible vertices can be extracted from higher-order correlation functions~\cite{weinberg1995quantum}. The standard procedure employs correlation functions involving large time differences. While this is very suitable for high-energy collider experiments, where an analysis is based on the concept of asymptotic states in the infinite past and future, this is not adequate for many realizations of strongly interacting many-body systems where the notion of an initial state `long before' and a final state `long after' the collision is not physical. Moreover, often these systems are studied at a given snapshot in time, without any direct reference to states in the asymptotic past or future. This is especially true for cold-atom experiments where one takes pictures, for example measuring every atom either after time of flight~\cite{Buecker2009} or in-situ~\cite{Bakr2009,Sherson2010}.
 
In this paper, we develop a pathway to extract the irreducible vertices of a quantum many-body system from experimental measurements. Our approach employs a formulation of QFT based on equal-time correlation functions only~\cite{wetterich1997nonequilibrium,nachbagauer1997wigner}. Equal-time correlation functions can be extracted from snapshot measurements~\cite{schweigler2017experimental,Feng2019,Rispoli2019} and, therefore, match well with experimental capabilities. We lay out the theoretical foundations of this approach, and illustrate the derivations using the sine-Gordon model. The irreducible vertices at equal times are estimated for this model both analytically and using numerical simulations. In particular, we show how to recover from the vertices the effective Hamiltonian underlying the dynamics. These theoretical results provide the basis for the benchmark verification of the QFT description extracted from experimental measurement. In the experiment, the sine-Gordon model is quantum simulated with two tunnel-coupled superfluids in thermal equilibrium~\cite{schweigler2017experimental}. We show how to extract the irreducible vertices from the experimental setup and compare the measurements to the theoretical estimates. The agreement of the experimental results with the theoretical expectations within errors provide a proof-of-principle verification of the approach. This represents an important step towards quantum simulator applications that are beyond reach of classical computational techniques. A first example of such an application is the recent experimental extraction of the irreducible two- and four-vertices for a strongly correlated spin-1 Bose condensate far from equilibrium~\cite{prufer2019experimental}, where no theoretical solution is available and which has been performed in parallel to this work.

The paper is organised as follows. 
We start in section \ref{sec:ET_QFT} with a self-contained description of an equal-time formulation of quantum field theory and  equal-time correlation functions as they arise naturally in experiments. In particular, we show how the one-particle irreducible (1PI) vertices, which constitute the fundamental building blocks of the QFT description of the many-body system, can be extracted from the measured equal-time correlation functions. 
In section \ref{sec:SG_thermal}, we illustrate these theoretical foundations in the framework of the quantum sine-Gordon (SG) model~\cite{Coleman75,Mandelstam,Faddeev19781,Sklyanin1979}  and calculate the 1PI correlation functions and the effective action in the classical field theory limit in thermal equilibrium and compare it to numerical simulations. 
As a proof-of-principle, we show in section \ref{sec:exp} an application to an experiment with two tunnel-coupled superfluids, which realises the SG model ~\cite{Gritsev2007,schweigler2017experimental}. 
We conclude our work in section \ref{sec:conlusion}.
An extensive appendix contains detailed calculations.

\section{\label{sec:ET_QFT}Extracting the irreducible vertices  from equal-time correlations }

In the standard formulation of quantum field theory one starts from a typical scattering experiment which gives access to the transition amplitude between an initial state at times long before the collision and its final state at much later times. These transition amplitudes determine the S-matrix elements, which can be expressed in terms of time-ordered correlation functions of the underlying quantum field theory~\cite{weinberg1995quantum}. Knowledge of all time-ordered correlation functions is then equivalent to solving the quantum theory~\cite{Schwinger1951A,Schwinger1951B}.

However, time-ordered correlation functions and the description by an S-matrix formulation are  conceptually less suitable in the analysis of strongly correlated complex quantum systems, which are often studied at a given snapshot in time. Such measurements at a given instant of time lead to the notion of equal-time correlation functions. In quantum field theory, these can be represented by expectation values of Weyl ordered products of field operators~\cite{cahill1969density,wetterich1997nonequilibrium}. Knowledge of all equal-time correlation functions at a given time $t$ contains all information about the many-body system at this instant of time. For example, the factorisation properties of higher-order correlation functions directly reveal if the system is free (factorising) or interacting (non factorising)~\cite{schweigler2017experimental}.  To extract the interaction constants of the underlying (effective) Hamiltonian one has to extract the so-called one-particle irreducible (1PI) correlation  functions~\cite{weinberg1995quantum} which represent the full non-perturbative interaction vertices of the quantum system.

While there  are standard textbook concepts to extract the 1PI correlation functions from time-ordered correlation functions, the possibility to extract them from equal-time correlation functions is much less explored. Here we illustrate how to extract them from the equal-time correlations and thereby show how to determine the effective Hamiltonian from experiment at a snapshot in time.

We start in section \ref{sec:ET_QFTa} with an introduction to quantum field theory in an equal-time formalism. At the example of a scalar field theory, we show the relation to Wigner's phase-space formalism commonly used e.g.~in quantum optics. We further summarise how to extract connected correlation functions (\ref{sec:ET_QFTb}) and one-particle irreducible vertices (\ref{sec:ET_QFTc}) by introducing suitable generating functionals. Finally, we approximately calculate the 1PI effective action in thermal equilibrium in section \ref{sec:ET_QFTd}, which provides a direct connection to the parameters of the microscopic Hamiltonian, and give a recipe on how to proceed (\ref{sec:ET_QFTe}).

\subsection{\label{sec:ET_QFTa}Equal-time formulation of quantum field theory}

The use of equal-time correlations is motivated by the progress of cold atomic setups which nowadays allow to extract highly resolved images at a given instant in time. It has long been known that QFT can be set up by only employing such equal-time information, without relying on multi-time correlations ~\cite{wetterich1997nonequilibrium,nachbagauer1997wigner}. This formulation has, however, never been widely used.  Theoretical progress in solving the equal-time formalism is  hampered by the lack of appropriate approximation schemes. Nevertheless, an equal-time formulation is perfectly suited to extract the irreducible vertices from experimental data representing a snapshot of the system at a fixed time.

Setting up an equal-time formulation relies on measurements of conjugate elementary operators that are non-commuting. As a consequence, one has to choose an ordering prescription~\cite{cahill1969density}. Since correlations in cold atom systems are straightforwardly obtained by multiplying and averaging single shot results, the obtained correlations correspond to a fully symmetrised (so-called Weyl) ordering of the quantum operators. Moreover, as we show below, this choice of ordering leads to a definition of 1PI correlators that is directly related to Hamiltonian parameters. For the rest of this paper, we thus focus on Weyl-ordered correlation functions. A short discussion of other ordering prescriptions is given in the appendix~\ref{app:ordering}.

For simplicity, we start with a real scalar field theory with Schrödinger field operators $\hat{\Phi}(\mathbf{x})$ and $\hat{\Pi}(\mathbf{x})$ that fulfill the canonical commuation relation
\begin{align}
\left[\hat{\Phi}(\mathbf{x}), \hat{\Pi}(\mathbf{y})\right] = i \hbar \delta\left(\mathbf{x}- \mathbf{y}\right) \; .
\end{align}
A general quantum state at time $t$ is in the Schrödinger picture described by the density operator $\hat{\rho}_t$.
Equivalently, knowing all correlations characterises the state $\hat{\rho}_t$ (see appendix~\ref{app:rho_from_correlations} for more details). Formally, all correlations can be conveniently summarised in the generating functional
\begin{align}\label{eq:Zt}
Z_t[J] =  \text{Tr} \left[ \hat{\rho}_t \exp \left(J^\varphi_\mathbf{x} \hat{\Phi}_\mathbf{x} + J^\pi_\mathbf{x} \hat{\Pi}_\mathbf{x} \right)\right] \; .
\end{align}
Here we have introduced a notation where repeated indices are integrated over, e.g.~$J^\varphi_\mathbf{x} \hat{\Phi}_\mathbf{x} = \int d^dx \, J^\varphi(\mathbf{x}) \hat{\Phi}(\mathbf{x})$. The $J$'s are so-called source fields, i.e.~auxiliary variables that encode the dependence of $\hat{\rho}_t$ on $\hat{\Phi}$ and $\hat{\Pi}$, as indicated by $\varphi$ and  $\pi$.
In the definition of $Z_t$, we have implemented the choice of ordering by treating the conjugate fields $\hat{\Phi}$ and $\hat{\Pi}$ symmetrically. The resulting correlation functions, which are obtained taking functional derivatives are Weyl-ordered. For example at second order, we have
\begin{align}
	G^{(2)}_{\mathbf{x},\mathbf{y}}(t) = \begin{pmatrix}
	\left\langle\varphi_{\mathbf{x}} \varphi_{\mathbf{y}} \right\rangle_{W_t} & \left\langle\varphi_{\mathbf{x}} \pi_{\mathbf{y}} \right\rangle_{W_t} \\
	\left\langle\pi_{\mathbf{x}} \varphi_{\mathbf{y}} \right\rangle_{W_t} & \left\langle\pi_{\mathbf{x}} \pi_{\mathbf{y}} \right\rangle_{W_t}
	\end{pmatrix} \;,
\end{align}
which consists of the three independent correlators
\begin{subequations}
	\begin{align}	
	\left\langle\varphi_{\mathbf{x}} \varphi_{\mathbf{y}} \right\rangle_{W_t} &= \left.\frac{\delta^2 Z_t[J]}{\delta J^\varphi_\mathbf{x} \delta J^\varphi_\mathbf{y}}\right|_{J=0} = \frac{1}{2}\text{Tr} \left[ \hat{\rho}_t \left(\hat{\Phi}_\mathbf{x} \hat{\Phi}_\mathbf{y} + \hat{\Phi}_\mathbf{y} \hat{\Phi}_\mathbf{x} \right)  \right] \;,\\
	\left\langle\varphi_{\mathbf{x}} \pi_{\mathbf{y}} \right\rangle_{W_t} &= \left.\frac{\delta^2 Z_t[J]}{\delta J^\varphi_\mathbf{x} \delta J^\pi_\mathbf{y}}\right|_{J=0} = \frac{1}{2}\text{Tr} \left[ \hat{\rho}_t \left(\hat{\Phi}_\mathbf{x} \hat{\Pi}_\mathbf{y} + \hat{\Pi}_\mathbf{y} \hat{\Phi}_\mathbf{x} \right)  \right] \;,\\
		\left\langle\pi_{\mathbf{x}} \pi_{\mathbf{y}} \right\rangle_{W_t} &= \left.\frac{\delta^2 Z_t[J]}{\delta J^\pi_\mathbf{x} \delta J^\pi_\mathbf{y}}\right|_{J=0} = \frac{1}{2}\text{Tr} \left[ \hat{\rho}_t \left(\hat{\Pi}_\mathbf{x} \hat{\Pi}_\mathbf{y} + \hat{\Pi}_\mathbf{y} \hat{\Pi}_\mathbf{x} \right)  \right] \;.
	\end{align}
\end{subequations}
This is explicitly verified in appendix ~\ref{app:ordering}, where also the higher-order case is discussed. 

For a general quantum many-body system, $Z_t[J]$ may involve more than one pair of canonically conjugated fields. These can be incorporated by adding appropriate sources $J$ and essentially does not affect the general discussion.

In the following, we consider only correlators of $\varphi$ to lighten the notation. Nevertheless, $\varphi$ may stand for either of the two fields and $\pi$ is only written explicitly when necessary to avoid confusion. 
In general, we then denote all Weyl-ordered correlators as
\begin{align}\label{eq:WeylCorrs}
G^{(n)}_{\mathbf{x}_1, \dots,\mathbf{x}_n}(t) = \left\langle\varphi_{\mathbf{x}_1} \cdots \varphi_{\mathbf{x}_n} \right\rangle_{W_t} = \left.\frac{\delta^{n} Z_t[J]}{\delta J_{\mathbf{x}_1} \cdots \delta J_{\mathbf{x}_n}}\right|_{J=0} \;.
\end{align}
We refer to the appendix~\ref{app:notation_summary} for a summary of all notational conventions used throughout this paper. In Eq.~\eqref{eq:WeylCorrs}, we have assumed a proper normalisation, $\Tr \hat{\rho}_t = 1$, which implies $Z_t[0] = 1$.

To make use of established, powerful QFT tools, we seek a representation of $Z_t$ in terms of functional integrals.
As shown in the appendix \ref{app:Zt_with_Wt}, the expression \eqref{eq:Zt} can be rewritten as a
\begin{align}\label{eq:Zt_with_Wt}
Z_t[J] = \int \mathcal{D}\varphi \, \mathcal{D}\pi\, 
W_t\left[\varphi,\pi\right] \exp \left(J^\varphi_\mathbf{x} \varphi_\mathbf{x}+J^\pi_\mathbf{x} \pi_\mathbf{x}\right) \;.
\end{align}
The integration kernel $W_t$ can be interpreted as a quasi-probability distribution,
\begin{align}\label{eq:Wigner_functional}
W_t[\varphi, \pi] = \int \mathcal{D}\tilde{\varphi} \left\langle \varphi - \frac{\tilde{\varphi}}{2} \right| \hat{\rho}_t\left| \varphi + \frac{\tilde{\varphi}}{2} \right\rangle \exp\left(\frac{i}{\hbar} \tilde{\varphi}_\mathbf{x} \pi_\mathbf{x}\right) \;,
\end{align}
the so-called Wigner functional. Here $|\varphi\rangle$ denotes an eigenstate of the field operator $\hat{\Phi}$ with eigenvalue $\varphi$. The Wigner function itself has previously been applied successfully in the context of quantum optics~\cite{hillery1984distribution} and also plays a prominent role in the semi-classical description of non-equilibrium quantum dynamics~\cite{polkovnikov2010phase, aarts2002classical}. Eq.~\eqref{eq:Zt_with_Wt} is the basis of the equal-time formulation of QFT and allows us to apply established procedures in the following.

\subsection{Connected correlation functions}\label{sec:ET_QFTb}
In QFT (and analogously in classical probability theory) it is well known that the correlations encoded in $Z_t$ are largely redundant \cite{shiryaev2016probability,weinberg1995quantum}. The first step is the removal of additive redundancies, by the introduction of another generating functional,
\begin{align}\label{eq:logZ}
E_t[J] = \log Z_t[J] \;.
\end{align}

We denote the corresponding correlations, called connected correlators, as
\begin{align}\label{eq:ConnCorrs}
G^{(n)}_{c,\mathbf{x}_1, \dots, \mathbf{x}_n}(t) = \left.\frac{\delta^{n} E_t[J]}{\delta J_{\mathbf{x}_1} \cdots \delta J_{\mathbf{x}_n}}\right|_{J=0} \;.
\end{align}
Explicitly, as shown in the appendix~\ref{app:conn_from_full}, up to fourth order they are given by
\begin{widetext}
\vspace*{-6mm}
\begin{subequations}\label{eq:howto_connected}
	\begin{align}
	G^{(1)}_{c,\mathbf{x}_1} &= G^{(1)}_{\mathbf{x}_1}\;, \label{eq:connected_a}\\
	G^{(2)}_{c,\mathbf{x}_1,\mathbf{x}_2} &= G^{(2)}_{\mathbf{x}_1,\mathbf{x}_2} - G^{(1)}_{c,\mathbf{x}_1} G^{(1)}_{c,\mathbf{x}_2} \;,\label{eq:connected_b}\\
	G^{(3)}_{c,\mathbf{x}_1,\mathbf{x}_2,\mathbf{x}_3} &= G^{(3)}_{\mathbf{x}_1,\mathbf{x}_2,\mathbf{x}_3} - \left(G^{(2)}_{c,\mathbf{x}_1,\mathbf{x}_2} G^{(1)}_{c,\mathbf{x}_3}  + G^{(2)}_{c,\mathbf{x}_2,\mathbf{x}_3} G^{(1)}_{c,\mathbf{x}_1} + G^{(2)}_{c,\mathbf{x}_3,\mathbf{x}_1} G^{(1)}_{c,\mathbf{x}_2} \right) - G^{(1)}_{c,\mathbf{x}_1} G^{(1)}_{c,\mathbf{x}_2} G^{(1)}_{c,\mathbf{x}_3} \;,\label{eq:connected_c}\\
	G^{(4)}_{c,\mathbf{x}_1,\mathbf{x}_2,\mathbf{x}_3,\mathbf{x}_4} &= 
	G^{(4)}_{\mathbf{x}_1,\mathbf{x}_2,\mathbf{x}_3,\mathbf{x}_4} - \left( G^{(3)}_{c,\mathbf{x}_1,\mathbf{x}_2,\mathbf{x}_3}G^{(1)}_{c,\mathbf{x}_4} +  G^{(3)}_{c,\mathbf{x}_2,\mathbf{x}_3,\mathbf{x}_4}G^{(1)}_{c,\mathbf{x}_1} +  G^{(3)}_{c,\mathbf{x}_3,\mathbf{x}_4,\mathbf{x}_1}G^{(1)}_{c,\mathbf{x}_2} +  G^{(3)}_{c,\mathbf{x}_4,\mathbf{x}_1,\mathbf{x}_2}G^{(1)}_{c,\mathbf{x}_3} \right)\nonumber\\
	&- \left(G^{(2)}_{c,\mathbf{x}_1,\mathbf{x}_2}G^{(2)}_{c,\mathbf{x}_3,\mathbf{x}_4} + G^{(2)}_{c,\mathbf{x}_1,\mathbf{x}_3} G^{(2)}_{c,\mathbf{x}_2,\mathbf{x}_4} + G^{(2)}_{c,\mathbf{x}_1,\mathbf{x}_4}G^{(2)}_{c,\mathbf{x}_2,\mathbf{x}_3} + \right) - 
	\left( G^{(2)}_{c,\mathbf{x}_1,\mathbf{x}_2} G^{(1)}_{c,\mathbf{x}_3}G^{(1)}_{c,\mathbf{x}_4} + G^{(2)}_{c,\mathbf{x}_1,\mathbf{x}_3} G^{(1)}_{c,\mathbf{x}_2}G^{(1)}_{c,\mathbf{x}_4} \right. \nonumber\\
	&+ \left. G^{(2)}_{c,\mathbf{x}_1,\mathbf{x}_4} G^{(1)}_{c,\mathbf{x}_2}G^{(1)}_{c,\mathbf{x}_3} +G^{(2)}_{c,\mathbf{x}_2,\mathbf{x}_3} G^{(1)}_{c,\mathbf{x}_1}G^{(1)}_{c,\mathbf{x}_4} +G^{(2)}_{c,\mathbf{x}_2,\mathbf{x}_4} G^{(1)}_{c,\mathbf{x}_1}G^{(1)}_{c,\mathbf{x}_3} +G^{(2)}_{c,\mathbf{x}_3,\mathbf{x}_4} G^{(1)}_{c,\mathbf{x}_1}G^{(1)}_{c,\mathbf{x}_2} \right) \nonumber\\
	&- G^{(1)}_{c,\mathbf{x}_1}G^{(1)}_{c,\mathbf{x}_2}G^{(1)}_{c,\mathbf{x}_3}G^{(1)}_{c,\mathbf{x}_4} \;,\label{eq:connected_d}
	\end{align}
\end{subequations}
\vspace*{-6mm}
\end{widetext}
where we suppressed the overall time-dependence for brevity.
For every order $n$, the connected part $G^{(n)}_c$, is obtained by subtracting the information already given by lower-order functions $G^{(m<n)}_c$.

\begin{figure}
	\centering\includegraphics[scale=0.16]{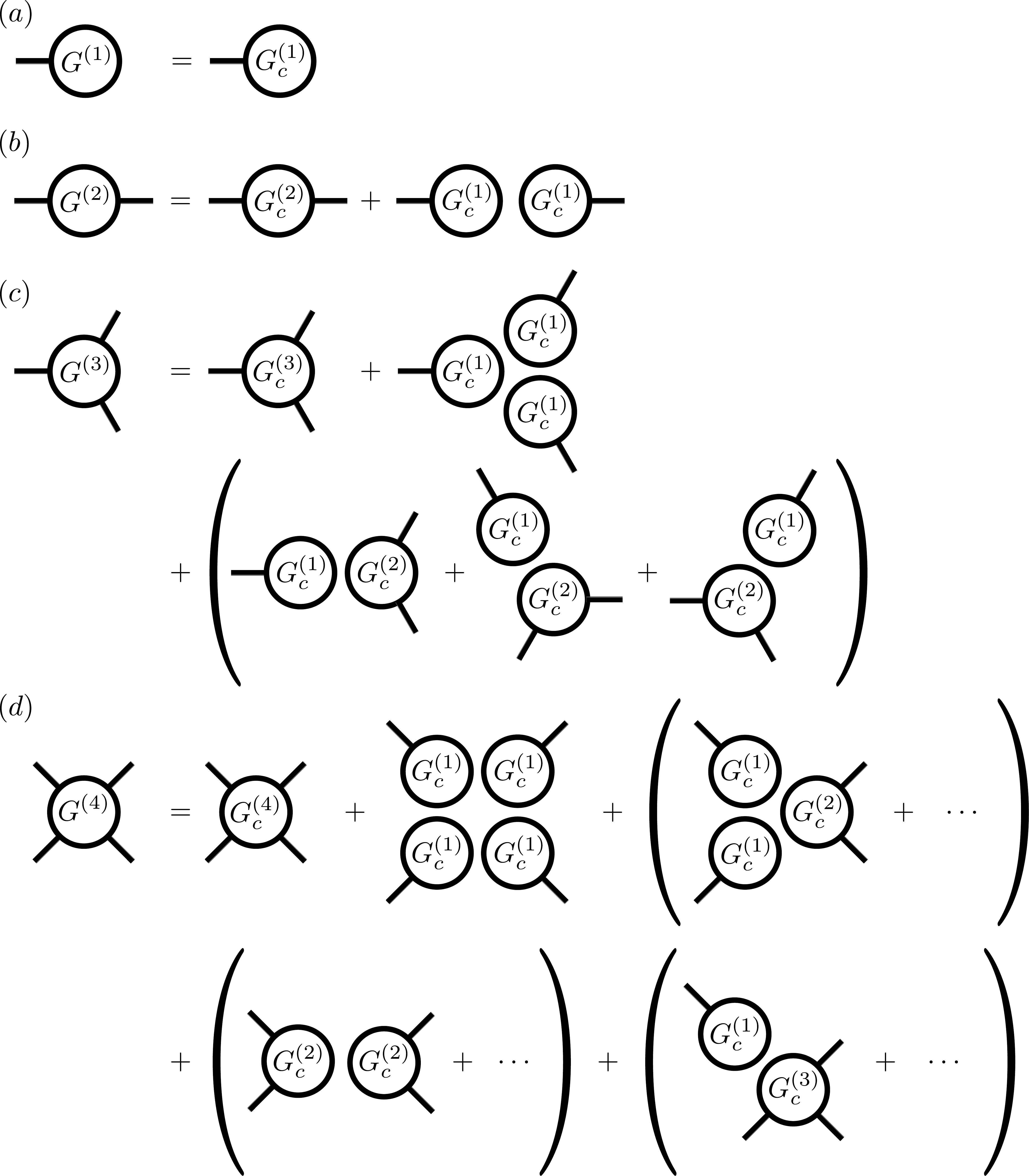}
	\caption{\label{fig:connected} {\bf Feynman diagrams relating full and connected correlation functions.} At first-order $(a)$ the correlations are identical. At second-order $(b)$ there is one disconnected diagram that contains redundant information. At higher orders an increasing number of disconnected diagrams need to be considered. We explicitly show the third-order $(c)$ and fourth-order $(d)$ correlations. The dots indicate permutations of the diagrammatic structure within the brackets, similar to $(c)$. }
\end{figure}

This can be visualised by a graphical representation in terms of Feynman diagrams, which is very helpful to organise the underlying combinatorics. By careful examination of this reorganisation, exemplified in Fig.~\ref{fig:connected}, one learns that only connected graphs contribute to the correlators generated by $E_t$, hence the name connected correlations. In short, taking the logarithm of $Z_t$ in Eq.~\eqref{eq:logZ} removes all disconnected diagrams.

Physically, this means that by inspecting the factorisation of higher-order correlation functions, one can determine whether or not the quantum system is described by a gaussian density operator $\hat{\rho}_t$. Since gaussian distributions correspond to free (non-interacting) QFTs, this in principle allows to determine the basis of conjugate fields which diagonalises the quantum many-body Hamiltonian that governs the system at hand.

\subsection{One-particle irreducible vertices}\label{sec:ET_QFTc}
The connected correlators of order higher than two still contain redundant information. In order to access the irreducible vertices, we define the effective action
\begin{align}\label{eq:eff_action}
	\Gamma_t[\Phi] = -E_t[J(\Phi)] + J_\mathbf{x}(\Phi) \Phi_\mathbf{x}
\end{align}
as a Legendre transform of $E_t$. In Eq.~\eqref{eq:eff_action} the relation $\Phi_\mathbf{x}(J) = \left(\delta E_t[J]\right)/\left(\delta J_\mathbf{x}\right)$ thus has to be inverted to obtain $J_\mathbf{x}(\Phi)$. We emphasise that the above notation is an abbreviation for a double Legendre transform in both of the conjugate fields $\Phi$ and $\Pi$. Accordingly, equations in this section implicitly include appropriate sums over the two fields.

Expanding the effective action in a functional Taylor series we have
\begin{align}\label{eq:taylor_Gamma}
	\Gamma_t[\Phi] = \sum_{n=2}^\infty \frac{1}{n!} \Gamma^{(n)}_{\mathbf{x}_1, \dots,\mathbf{x}_n}(t) \prod_{j=1}^n \left(\Phi_{\mathbf{x}_j} -  \bar{\Phi}_{\mathbf{x}_j}(t)\right)\; .
\end{align}
Here $\bar{\Phi}_\mathbf{x}(t) = \langle \varphi_\mathbf{x} \rangle_{W_t}$ is the mean value at time $t$, for which the effective action is stationary, i.e.~$\left(\delta \Gamma_t[\Phi]\right)/\left(\delta \Phi\right)|_{\Phi = \bar{\Phi}} = 0$.
The 1PI vertices,
\begin{align}\label{eq:1PICorrs}
\Gamma^{(n)}_{\mathbf{x}_1, \dots, \mathbf{x}_n}(t) = \left.\frac{\delta^{n} \Gamma_t[\Phi]}{\delta \Phi_{\mathbf{x}_1} \cdots \delta \Phi_{\mathbf{x}_n}}\right|_{\Phi=\bar{\Phi}} \;,
\end{align}
are the expansion coefficients in this series.
In Eq.~\eqref{eq:taylor_Gamma}, the sum starts at $n=2$ because we have omitted an irrelevant constant $\Gamma^{(0)}$ and the first order, $\Gamma^{(1)}$, vanishes by construction due to the expansion around $\bar{\Phi}$. Physically, $\bar{\Phi}$ can take a non-vanishing value, which plays a crucial role, e.g., in the case of spontaneous symmetry breaking or the false vacuum decay~\cite{Coleman77A}.

\begin{figure}
	\centering{\includegraphics[scale=0.16]{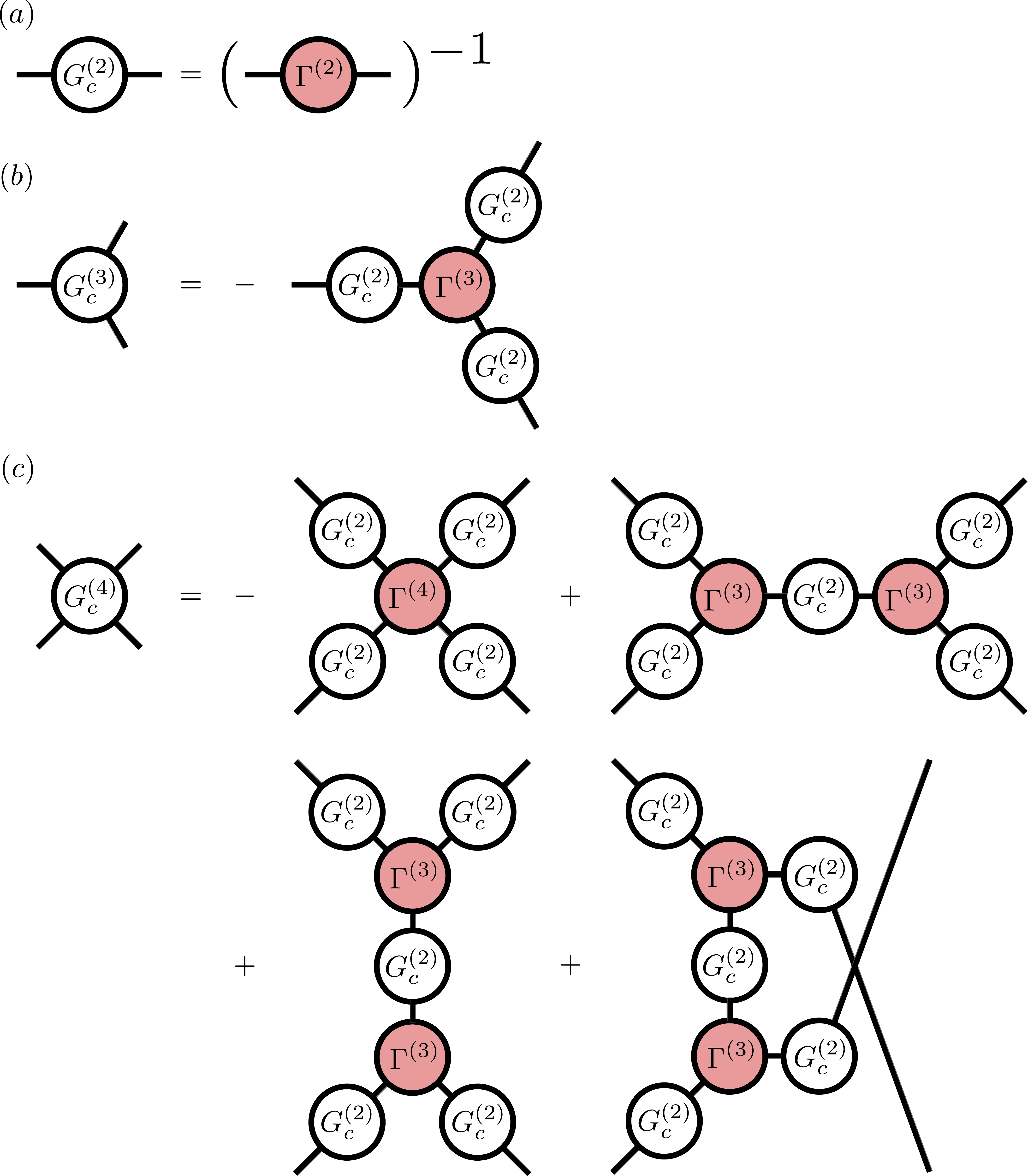}}
	\caption{\label{fig:1PI}{\bf Feynman diagrams relating connected and 1PI correlation functions.} At second order $(a)$, the correlators are each others inverse. At third order $(b)$, the connected correlator is `built' from connected two-point functions that are connected by the irreducible three-vertex. At fourth order $(c)$, the structure is similar to $(b)$ with contributions from the four- and three-vertices. Given the 1PI vertices $\Gamma^{(n)}$, all connected correlations can be calculated by summing so called tree-diagrams, which separate into two disconnected parts upon cutting a single line $G^{(2)}_c$ and hence do not contain any loop diagrams.}
\end{figure}

As shown in the appendix \ref{app:1PI_from_conn}, the 1PI vertices up to fourth order are related to the connected correlation functions as follows:
\begin{widetext}
\vspace*{-6mm}
	\begin{subequations}\label{eq:explicit1PI}
		\begin{align}
		\Gamma^{(1)}_{\mathbf{x}_1} &= 0\;, \\
		\Gamma^{(2)}_{\mathbf{x}_1,\mathbf{x}_2} &=  \left[G_c^{(2)}\right]^{-1}_{\mathbf{x}_1,\mathbf{x}_2} \;, \label{eq:1PI_b}\\
		\Gamma^{(3)}_{\mathbf{x}_1,\mathbf{x}_2,\mathbf{x}_3} &= -\Gamma^{(2)}_{\mathbf{x}_1,\mathbf{y}_1} \Gamma^{(2)}_{\mathbf{x}_2,\mathbf{y}_2} \Gamma^{(2)}_{\mathbf{x}_3,\mathbf{y}_3}G^{(3)}_{c,\mathbf{y}_1,\mathbf{y}_2,\mathbf{y}_3} \;, \label{eq:1PI_c}\\
		\Gamma^{(4)}_{\mathbf{x}_1,\mathbf{x}_2,\mathbf{x}_3,\mathbf{x}_4} &= - \Gamma_{\mathbf{x}_1,\mathbf{y}_1} \Gamma_{\mathbf{x}_2,\mathbf{y}_2} \Gamma_{\mathbf{x}_3,\mathbf{y}_3} \Gamma_{\mathbf{x}_4,\mathbf{y}_4} G^{(4)}_{c,\mathbf{y}_1,\mathbf{y}_2,\mathbf{y}_3,\mathbf{y}_4} \nonumber\\
		&+ \Gamma^{(2)}_{\mathbf{x}_1,\mathbf{y}_1} \Gamma^{(2)}_{\mathbf{x}_2,\mathbf{y}_2} \Gamma^{(2)}_{\mathbf{x}_3,\mathbf{y}_3} \Gamma^{(2)}_{\mathbf{x}_4,\mathbf{y}_4}
		 \Gamma^{(2)}_{\mathbf{z}_1,\mathbf{z}_2}
		 \left(G^{(3)}_{c,\mathbf{y}_1,\mathbf{y}_2,\mathbf{z}_1} G^{(3)}_{c,\mathbf{z}_2,\mathbf{y}_3,\mathbf{y}_4}+ G^{(3)}_{c,\mathbf{y}_1,\mathbf{y}_3,\mathbf{z}_1} G^{(3)}_{c,\mathbf{z}_2,\mathbf{y}_2,\mathbf{y}_4}  + G^{(3)}_{c,\mathbf{y}_1,\mathbf{y}_4,\mathbf{z}_1} G^{(3)}_{c,\mathbf{z}_2,\mathbf{y}_2,\mathbf{y}_3}  \right) \;. \label{eq:1PI_d}
		\end{align}
	\end{subequations}
\end{widetext}
We again emphasise that the explicit equations should be understood including appropriate sums over $\varphi$ and $\pi$ correlators (see appendix~\ref{app:1PI_from_conn}).
For higher orders these relations become more complicated and calculations are conveniently performed with the graphical notation exemplified in Fig.~\ref{fig:1PI}. These diagrams also explain the attribute 1PI: The diagrams representing the vertices can not be disconnected by cutting a single line. In this sense they are the irreducible structures from which all correlation functions and thus all physical observables can be recovered.

This also justifies the name effective action: $\Gamma_t$ is the quantum generalisation of a classical action including all corrections due to quantum-statistical fluctuations. However, in contrast to the `standard' (unequal-time) action, there is one stationarity condition for each time $t$. Together, they do not give a time evolution equation for the one-point function in the usual sense, but one differential equation for each time $t$. In that sense the time $t$ is treated as a label in the equal-time formulation of QFT.

\subsection{Measuring the effective Hamiltonian}\label{sec:ET_QFTd}

So far, we have equivalently rewritten the quantum-statistical information of a system described by a density operator $\hat{\rho}_t$ in terms of generating functionals $Z_t$, $E_t$ and $\Gamma_t$ which encode full, connected and 1PI correlation functions, respectively. While the entries of the density matrix are typically inaccessible and less intuitive, the equal-time correlators can be measured in experiments and are directly related to relevant observables and structural information, such as occupation numbers and couplings.  Next, we employ the equal-time formalism to relate parameters of an Hamiltonian to the 1PI correlators.

As a generic example we consider a relativistic scalar field theory with potential $V$ described by the Hamiltonian
\begin{align}
\hat{H} =\int_\mathbf{x} \left[\frac{1}{2}\,\hat{\Pi}_\mathbf{x}^2 + \frac{1}{2}\left(\nabla_\mathbf{x}\hat{\Phi}_\mathbf{x}\right)^2  + V \left(\hat{\Phi}_\mathbf{x}\right)\right] \; .
\end{align}	
Given the Hamiltonian $\hat{H}$, it is possible to derive an evolution equation for $\Gamma_t$~\cite{wetterich1997nonequilibrium}.
For simplicity, we focus on the case of thermal equilibrium, which is a stationary solution $\Gamma_\beta$, described by the density operator $\hat{\rho}_\beta \sim \exp \left(- \beta \hat{H} \right)$ with the prefactor fixed by normalisation.  

In order to obtain the generating functional Eq.~\eqref{eq:Zt_with_Wt} we need to calculate the Wigner functional Eq.~\eqref{eq:Wigner_functional}. In the interacting case, the involved functional integration can only be performed approximately. It is however possible to derive an exact equation for the thermal Wigner functional (see appendix~\ref{app:W_thermal}). It takes the form of a functional flow equation, $\partial_\beta W_\beta = - (H_0 + \hbar^2 H_1 + \dots )W_\beta$, with
	\begin{align}
	H_0 &= \int_\mathbf{x} \left[\frac{1}{2} \pi_\mathbf{x}^2 + \frac{1}{2} \left(\nabla_\mathbf{x}\varphi_\mathbf{x}\right)^2 +V_\mathbf{x}\left(\varphi\right)\right] \; .
	\end{align}
It is straightforward to solve the equation for $W_\beta$ perturbatively by a semi-classical expansion in powers of $\hbar$. 

The leading order is the classical field theory limit, where we obtain $W_\beta \sim \exp \left(-\beta H_0\right)$ with the classical Hamiltonian $H_0$. Then the generating functional Eq.~\eqref{eq:Zt_with_Wt} becomes
\begin{align}\label{eq:Z_thermal}
Z_\beta[J] \sim \int \mathcal{D}\pi \mathcal{D}\varphi \; e^{-\beta H_0 + J^\varphi_\mathbf{x}\varphi_\mathbf{x} + J^\pi_\mathbf{x}\pi_\mathbf{x}} \; .
\end{align}
Thus $\beta H_0$ plays the role of a classical action for the fluctuating fields $\varphi$ and $\pi$. This allows us to calculate the effective action $\Gamma_\beta$ in the equal-time formalism using established QFT methods, such as a the background field method employed below. 

We note that the two conjugate fields $\varphi$ and $\pi$ decouple in the present limit, which implies that the effective action separates as
\begin{align}\label{eq:separation_of_phi_and_pi}
\Gamma_\beta[\Phi,\Pi] = \Gamma_\beta[\Phi] + \Gamma_\beta [\Pi] \;
\end{align}
with $\Gamma_\beta[\Pi] = \frac{\beta}{2} \int_\mathbf{x}  \Pi_\mathbf{x}^2 + \text{const.}$, as shown in appendix~\ref{app:separation_phi_and_pi}.
The separation Eq.~\eqref{eq:separation_of_phi_and_pi} is a property of the classical field approximation. In general, the quantum effective action of the full quantum theory requires knowledge of all equal-time correlators of $\varphi$ and $\pi$, including mixed terms. However, symmetries such as time translation invariance simplify the discussion, see appendix~\ref{app:B8}.

By means of the background field method, we can calculate the effective action in a loop expansion. In the present formalism, we split
\begin{align} \label{eq:measure_H}
\Gamma_\beta[\Phi] = \beta H[\Phi] + \Gamma'_\beta[\Phi] \;,
\end{align}
where $H[\Phi] = H_0[\varphi = \Phi, \pi = 0]$. As shown in the appendix~\ref{app:derivation_loop_expansion}, the `rest' $\Gamma_\beta'$ obeys the following functional integro-differential equation,
\begin{align}\label{eq:Gamma_eq}
e^{-\Gamma'_\beta[\Phi]} = \int \mathcal{D}\varphi \; \exp \left(-\beta K[\varphi,\Phi] + \frac{\delta \Gamma'_\beta[\Phi]}{\delta \Phi_\mathbf{x}}\varphi_\mathbf{x}\right)\;,
\end{align}
where we abbreviated
	\begin{align}
	K[\varphi,\Phi] &= H[\Phi + \varphi] - H[\Phi] - \int_\mathbf{x} \frac{\delta H[\Phi]}{\delta \Phi_\mathbf{x}}\;\varphi_\mathbf{x}\; .
	\end{align}
The solution of this equation is organised diagrammatically as an expansion in the number of loops. At leading order (tree-level) in this expansion $\Gamma'_\beta = 0$ and thus the equal-time effective action is directly related to the microscopic Hamiltonian. Consequently, the 1PI vertices correspond to the interaction constants of the underlying system. Beyond the leading-order approximation, the notion of the microscopic Hamiltonian becomes a less useful concept. The effective action then plays the role of an effective Hamiltonian, with all corrections from quantum-statistical fluctuations taken into account.

Returning to the leading order approximation, which gives rise to the tree-level 1PI vertices, we explicitly have
\begin{subequations}\label{eq:measure_vertices}
\begin{align}
\Gamma^{(2)}_{\mathbf{x},\mathbf{y}} &=  \nabla_\mathbf{x}^2\delta(\mathbf{x}-\mathbf{y})   +\int_\mathbf{z}\left. \frac{\delta^2 V_\mathbf{z}(\Phi)}{\delta \Phi_\mathbf{x}\delta \Phi_\mathbf{y}} \right|_{\Phi=\bar{\Phi}}\;,\\
\Gamma^{(n)}_{\mathbf{x_1}, \dots, \mathbf{x_{n}}} &= \int_\mathbf{z}\left. \frac{\delta^n V_\mathbf{z}(\Phi)}{\delta \Phi_{\mathbf{x}_1} \cdots \delta \Phi_{\mathbf{x}_n}} \right|_{\Phi=\bar{\Phi}}\;,
\end{align}
\end{subequations}
where $n\ge3$.
Eq.~\eqref{eq:measure_H} or more explicitly Eq.~\eqref{eq:measure_vertices} directly show the relation between the 1PI correlation functions and the parameters of the microscopic (or more generally an effective) Hamiltonian. 
Together with the procedure to obtain the 1PI correlators, outlined below, they provide an experimental prescription for measuring a quantum many-body Hamiltonian. 

\subsection{Recipe to extract 1PI correlators}\label{sec:ET_QFTe}

The extraction of 1PI correlators, which are the fundamental irreducible building blocks for the QFT description, from equal-time data proceeds in the following steps.  

\begin{enumerate}
	\item{Identify the degrees of freedom of interest which constitute the elementary fields $\varphi$ of the QFT. }
	\item{Obtain many realisations ($i = 1, \dots, N$) of the desired field $\varphi_i(\mathbf{x})$ at the times $t$ of interest.}
	\item{Estimate the full correlators up to order $n$ by averaging $G^{(n)}_{\mathbf{x}_1, \dots, \mathbf{x}_n } \approx \frac{1}{N} \sum_i \varphi_i(\mathbf{x}_1) \cdots \varphi_i(\mathbf{x}_n)$}.
	\item{Obtain the connected correlators $G_c^{(n)}$ by subtracting the disconnected contributions according to Eq.~\eqref{eq:howto_connected}}.
	\item{Calculate the 1PI correlators $\Gamma^{(n)}$ by reducing the connected correlators according to Eq.~\eqref{eq:explicit1PI}}.
\end{enumerate}

This procedure corresponds to a shift of representation from the density operator $\hat{\rho}$ to $\Gamma$, the generating functional for 1PI correlators.
In the following two sections we will illustrate and verify the method in the case of the sine-Gordon model with numerically simulated data (Section \ref{sec:SG_thermal}) and with experimental measurements (Section \ref{sec:exp}).

\section{Example: sine-Gordon model} \label{sec:SG_thermal}

As an explicit example, we consider the sine-Gordon model ~\cite{Coleman75,Mandelstam,Faddeev19781,Sklyanin1979}  in thermal equilibrium.  It is an interacting relativistic scalar field theory described by
\begin{align}
\beta \hat{H}_\text{SG} = \int_x \left\lbrace \beta g\hat{\Pi}_x^2 +  \frac{\lambda_T}{4}\left[  \frac{1}{2}\left(\partial_x\hat{\Phi}_x\right)^2 - \frac{1}{\ell_J^2}\cos\left( \hat{\Phi}_x\right)\right] \right\rbrace \; , \label{eq:H_SG}
\end{align}
where $\beta = (k_\mathrm{B} T)^{-1}$ is the inverse temperature. The specific form of the Hamiltonian $\hat{H}_\text{SG} $ given above is motivated by the recent progress to quantum simulate the SG model by two tunnel coupled 1D superfluids~\cite{Gritsev2007,schweigler2017experimental}. See section \ref{sec:exp} for the physical origin of the fields $\hat{\Phi}$ and $\hat{\Pi}$, the microscopic parameter $g$ and the length scales $\lambda_T$ and $\ell_J$. 

The semiclassical approximation Eq.~\eqref{eq:Z_thermal} is valid for 
\begin{align}\label{eq:validity_semiclass}
 \sqrt{4 \gamma} \ll \operatorname{min}\left[1,\frac{4}{Q}\right] \, ,
\end{align}
where the dimensionless parameters are $\gamma = 16 g \beta/\lambda_T$ and $Q = \lambda_T / \ell_J$. In the semi-classical limit the loop expansion is controlled by $Q$ with the tree-level approximation valid for $1/Q \ll 1$ (see appendix \ref{app:W_thermal} and \ref{app:derivation_loop_expansion} for details). We therefore consider in the following $\lambda_T = 17.35 \, \mu$m and vary $\ell_J$ such that $1 \lesssim Q \lesssim 20$.

Following the general discussion of the previous section, the tree-level vertices corresponding to Eq.~\eqref{eq:H_SG} are 
\begin{subequations}\label{eq:tree-level1PI}
	\begin{align}
	\Gamma^{(2),\text{tree}}_{p} &=  \frac{\lambda_T}{4}\left(p^2 + \frac{1}{\ell_J^2}\right) \;,\\
	\Gamma^{(2n),\text{tree}}_{p_1, \dots p_{2n-1}}&= - \frac{\lambda_T}{4\ell_J^2} \, (-1)^n \;, \\
	\Gamma^{(2n-1),\text{tree}}_{p_1, \dots p_{2n-2}}&= 0\;,
	\end{align}
\end{subequations}
where $n>2$ and we switched to momentum space correlators. Here and in the following we always consider the diagonal part in momentum space, i.e.~\mbox{$\Gamma^{(n)}_{p_1, \dots, p_n} = \left(2\pi\right) \delta (p_1 + \dots + p_n) \Gamma^{(n)}_{p_1, \dots, p_{n-1}}$}, which removes the volume factors arising from translation invariance. Note that $\Gamma^{(2n-1)} = 0 \; (\forall n \geq 1)$ remains valid beyond the tree-level approximation due to the symmetries of the SG Hamiltonian.

\begin{figure}
	\includegraphics[width=\columnwidth]{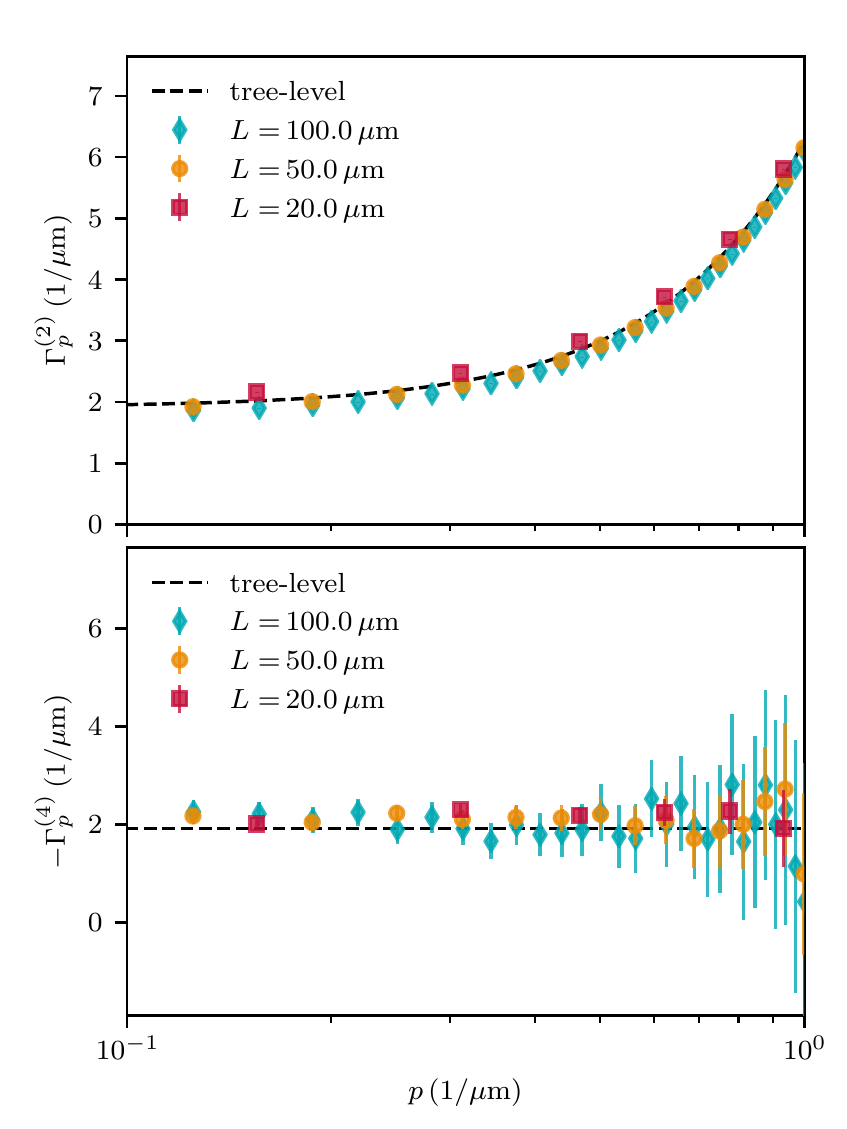}
	\caption{\label{fig:comparison_to_treelevel}{\bf 1PI vertices in the weak coupling regime} at $Q \approx 11.52$ calculated from the numerical data for different volumes $L=20 \mu$m (\redsquare), $50 \mu$m (\orangecircle), and $100 \mu$m (\bluediamond).
		The 1PI two-point function (upper panel) and 1PI 4-vertex (lower panel) show excellent agreement with the tree-level prediction (dashed black line) for a wide range of momenta. The results have been obtained from $10^8$ samples. The error bars indicate the standard error of the mean.
	}
\end{figure}

Employing a stochastic process based on a transfer matrix formalism \cite{Beck18}, we numerically obtain thermal profiles of the field $\varphi_x$, corresponding to the operator $\hat{\Phi}_x$. These are exact solutions of the SG model within the semi-classical approximation and hence have contributions up to arbitrary order in the above loop expansion. With these statistical samples, we carry out the procedure described in section~\ref{sec:ET_QFTe} and calculate the 1PI correlators up to fourth order.

So far, we have implicitly assumed in Eq.~\eqref{eq:tree-level1PI} that the correlations are obtained for an infinite system with periodic boundary conditions. The employed numerics, however, yield correlators from a finite subsystem, which is better described by open boundary conditions. We therefore employ a cosine transform and translate the results to momentum space, i.e.~Fourier momenta (for details see appendix~\ref{app:data_analysis}). 

Figure~\ref{fig:comparison_to_treelevel} shows the calculated 1PI vertices in momentum space for a large value of $Q \approx 11.5$ and different volumes $L$. Note that due to the periodicity of the sine-Gordon potential\footnote{The SG model is invariant under the shift $\varphi \to \varphi + \mathbb{Z} \times 2 \pi$. This leads to an undefined offset for the numerical profiles $\varphi_x$ and hence an undefined value of the momentum correlators for $p=0$.} the value of $p=0$ is not defined for the correlations considered. Therefore the fact that the correlation function is diagonal is crucial to be able to perform the inversion of the connected second order correlation function in order to obtain $\Gamma^\mathrm{(2)}$. The diagonal form allows to do the inversion for $p \neq 0$ without knowing the value for $p=0$.

We find excellent agreement with the tree-level predictions for the momentum diagonal of the two and four vertex. This demonstrates the possibility to carry out the procedure described in the previous section, which allows to directly measure the microscopic parameters through equal-time 1PI correlation functions if higher-order loop corrections can be neglected.

\begin{figure}
	\includegraphics[width=\columnwidth]{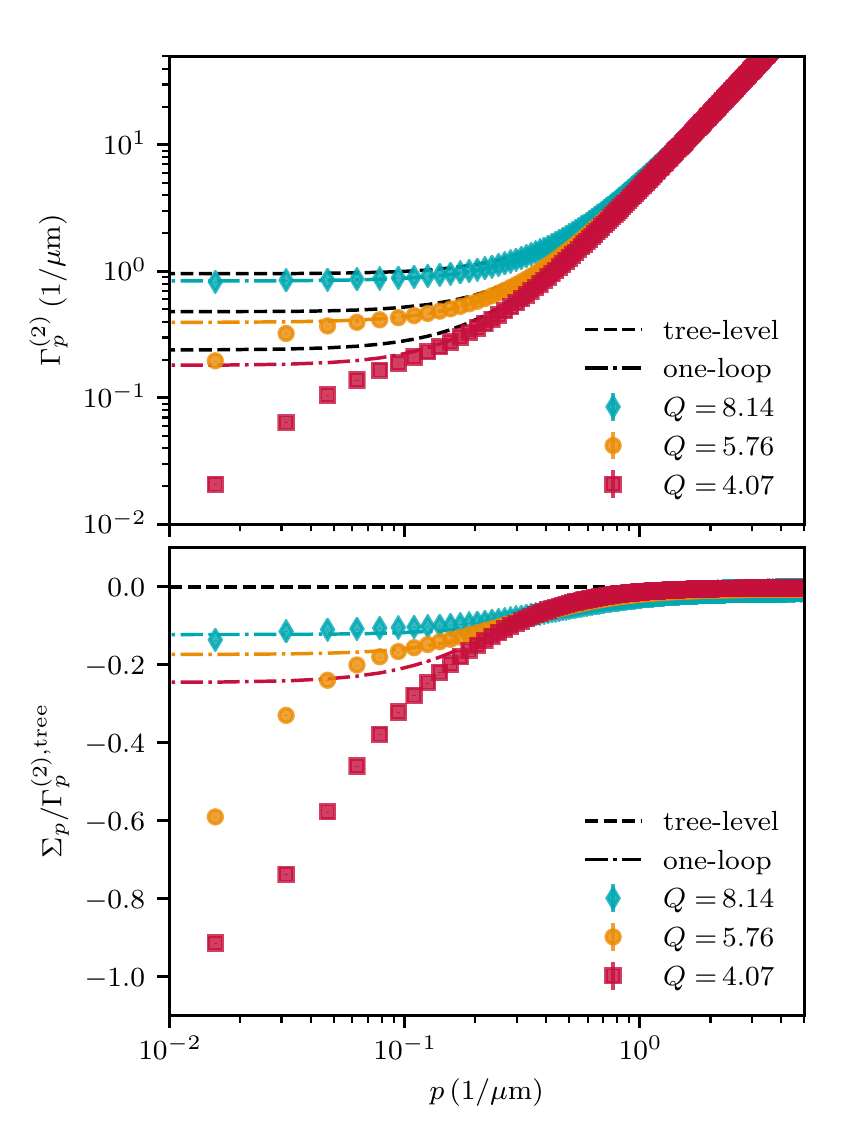}
	\caption{\label{fig:self_energy}{\bf Loop corrections to the 1PI two-point function} for different $Q$ in the strong coupling regime, $Q = 8.14$ (\bluediamond), $5.76$ (\orangecircle), and $4.07$ (\redsquare). The calculated two-point function (upper panel) always approaches the corresponding tree-level predictions (black dashed lines) at high momenta, as expected. In the infrared, we observe deviations due to loop corrections. The corresponding coloured dashed-dotted lines include the one-loop correction. The corresponding self-energy (lower panel) quantifies the deviations from the tree-level prediction. The corrections become more pronounced for smaller $Q$, as expected. Numerical results are calculated for $L = 200\, \mu\text{m}$ and a sample size of $10^8$.
	}
\end{figure}

However, when the system becomes strongly correlated the microscopic details become irrelevant and replaced by effective, momentum-dependent (so-called running) couplings. This behaviour is precisely captured by the 1PI correlation functions, which effectively replace the microscopic coupling parameters appearing in the Hamiltonian $H_\mathrm{SG}$. We therefore adjust the parameters away from the weakly-coupled limit $Q \gg 1$. In general, we observe an increasing deviation from the tree-level approximation, which is expected because loop corrections due to increasing fluctuations modify the physics.

Quantitatively, the corrections to the 1PI two-point function are summarized in the self-energy $\Sigma$,  defined via
\begin{align}
\Gamma^{(2)}_{p} = \Gamma^{(2),\text{tree}}_{p}+ \Sigma_p \;.
\end{align} 
In the appendix \ref{app:One-loop}, we calculate the leading correction, 
\begin{align}
	\Sigma_p^{\text{one-loop}} = -1/(4\ell_J) \; .
\end{align}

In Fig.~\ref{fig:self_energy}, the 1PI two-point function and the self-energy are plotted as a function of $p$. Generically, tree-level dominates in the ultraviolet (i.e.~at high momenta), which we also observe numerically. The 1PI two-point function approaches the power-law $\propto p^2$ in this limit and the (normalised) self-energy vanishes.

In the infrared (low momenta), however,  loop corrections are important. It is this regime where collective macroscopic phenomena emerge and the microscopic details are washed out. We observe a negative self-energy and hence a reduction of the 1PI two-point function, which agrees with the one-loop result over an intermediate range of momenta (and $Q$). Physically, this result implies stronger fluctuations as $Q$ decreases, consistent with the expectation for a strongly correlated regime of the sine-Gordon model.

\begin{figure}
	\includegraphics[width=\columnwidth]{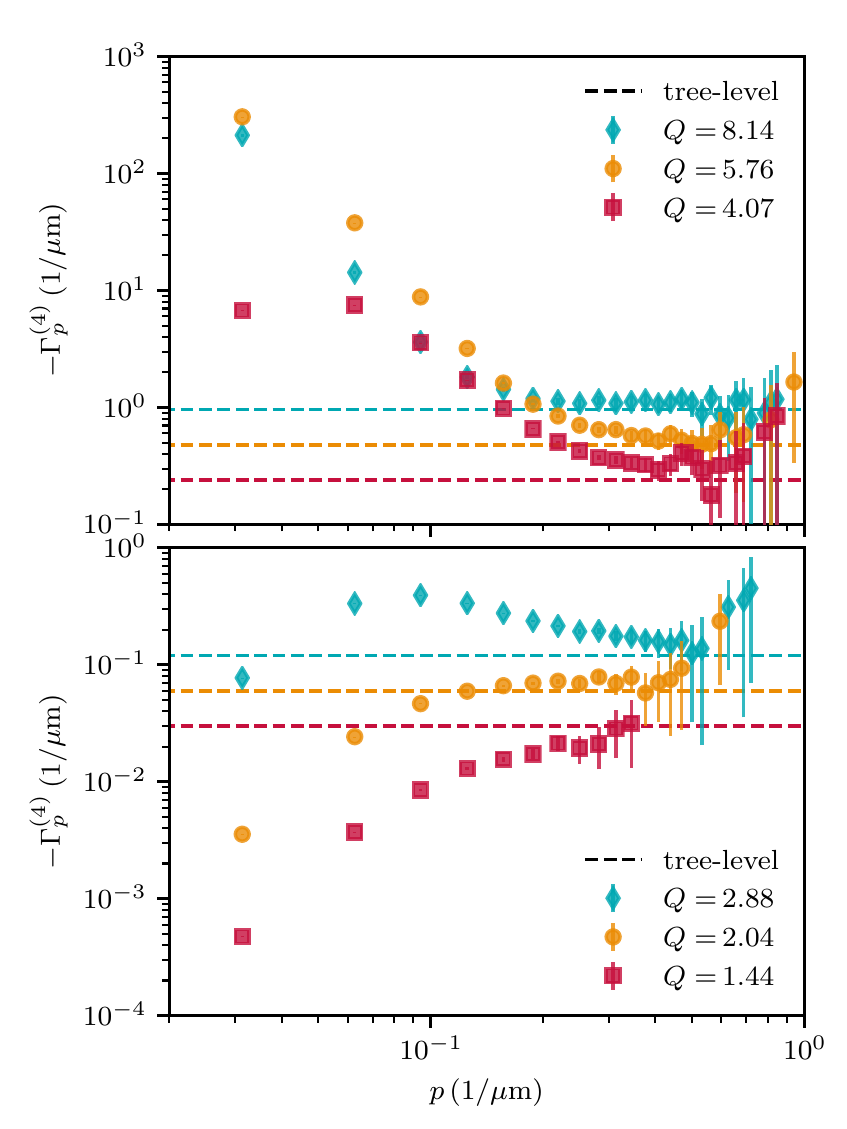}
	\caption{\label{fig:running_coupling}{\bf Loop corrections to the 1PI 4-vertex} for different values of $Q$. The (negative) 4-vertex (\bluediamond, \orangecircle, \redsquare), shown for larger values of $Q$ (upper panel, corresponding to Fig.~\ref{fig:self_energy}), clearly approaches the corresponding tree-level predictions (coloured dashed lines) at high momenta. In the infrared, we observe a strong momentum dependence, increasing the effective coupling. For decreasing values of $Q$ (lower panel) the (negative) 4-vertex is increasingly suppressed in the infrared. The results are consistent with the approach of the tree-level prediction (coloured dashed lines) for high momenta.
	Numerical results are calculated for $L = 100\, \mu\text{m}$ and a sample size of $10^8$. The error bars indicate the standard error of the mean. Note that we excluded data points at high momenta with errors larger than the mean from this plot.
	}
\end{figure}

Similarly, the loop corrections to the 1PI vertices lead to the notion of running couplings, i.e.~momentum-dependent interaction vertices that deviate from the constant microscopic values. In the appendix~\ref{app:One-loop}. we calculate the one-loop vertex
\begin{align}
	\Gamma^{(4),\text{one-loop}}_p = - \frac{\lambda_T}{4 \ell_J^2} - \frac{1}{8 \ell_J^3} \frac{1}{p^2 + 1/\ell_J^2} \; .
\end{align} 
Again, it is expected that loop corrections vanish for high momenta and the 1PI four-vertex converges to the tree-level result, i.e.~the microscopic parameters of the Hamiltonian, which is confirmed by our numerical simulations.

This is demonstrated in Fig.~\ref{fig:running_coupling}, where the 1PI four-vertex is shown for the same values of $Q$ as in Fig.~\ref{fig:self_energy}. For very high momenta, we are again limited by finite statistics. In the infrared, we clearly observe the momentum-dependent, i.e.~running, coupling. The increased values indicate stronger interactions, qualitatively consistent with the one-loop calculation. The effect is again more pronounced for smaller values of $Q$, as expected in the strongly correlated regime of the sine-Gordon model. 

\begin{figure}
	\includegraphics[width=\columnwidth]{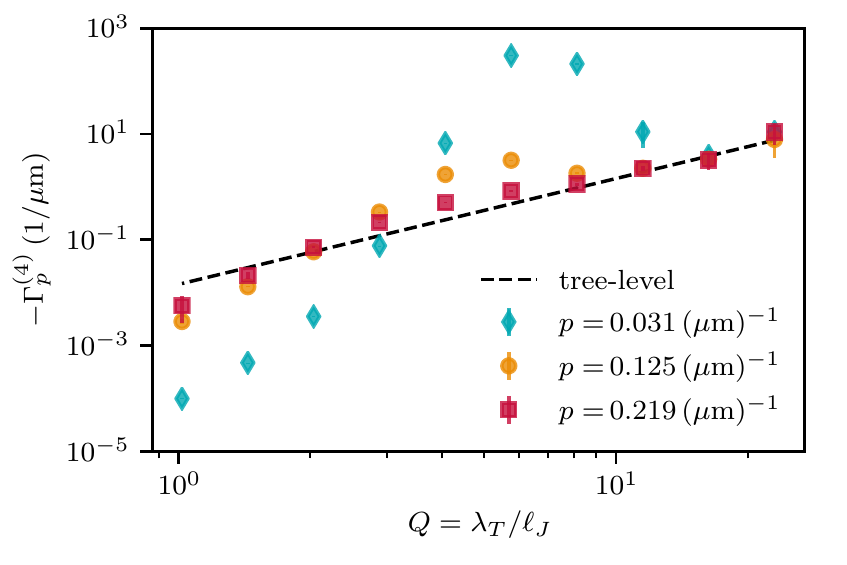}
	\caption{\label{fig:Q_dependence}{\bf 1PI 4-vertex as a function of $Q$} for three different momenta $p=0.031 \mu\mathrm{m}^{-1}$ (\bluediamond), $0.125 \mu\mathrm{m}^{-1}$ (\orangecircle), and $0.219 \mu\mathrm{m}^{-1}$ (\redsquare). At large momenta, the vertex approaches the tree-level prediction (black dashed-line). At low momenta, loop correction lead to a suppression or an enhancement depending on the values of $Q$ and $p$ (c.f.~Fig.~\ref{fig:running_coupling}). Numerical results are calculated for $L = 100\, \mu\text{m}$ and a sample size of $10^8$.
}
\end{figure}

We observe a qualitative difference between large and small values of $Q$. For $Q \gtrsim 4$, the magnitude of the 1PI four-vertex is increased in the infrared and shows a running coupling towards the smaller tree-level value. For $Q \lesssim 3$, the magnitude of the vertex decreases in the infrared as compared to the tree-level value at higher momenta. This behaviour is also clearly visible in Fig.~\ref{fig:Q_dependence}, where we show the four-vertex as a function of $Q$ for fixed momenta.

\section{Experimental results: Proof-of-principle}
\label{sec:exp}

\begin{figure}[b]
	\centering
	\includegraphics[width=\columnwidth]{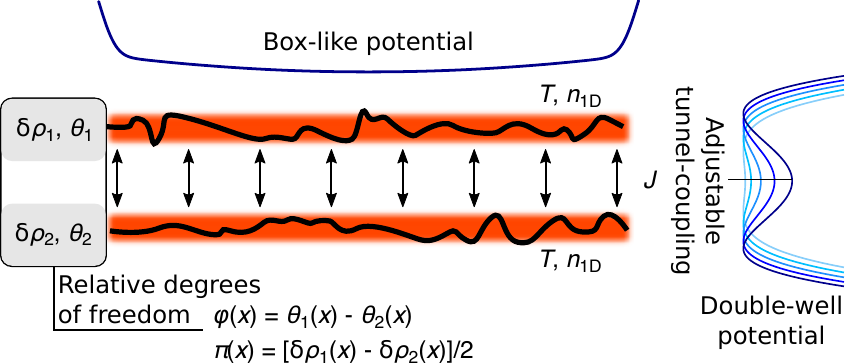}
	\caption{{\bf Schematics of the experimental setup.} We consider two tunnel-coupled one-dimensional superfluids in a double-well potential at a common temperature $T$. Changing the barrier height of the potential (blue lines) allows for an adjustable tunnel-coupling $J$ between the two superfluids. The superfluids are described in terms of density fluctuations $\delta\rho_{1,2}$ around their equal mean densities $n_{\mathrm{1D}}$ and fluctuating phases $\theta_{1,2}$ (black lines). From these quantities we define the relative degrees of freedom $\pi$ and $\varphi$ which represent the conjugate fields in the sine-Gordon Hamiltonian. 
	Figure adapted from \cite{schweigler2017experimental}.
	}
	\label{fig:Schematics}
\end{figure}

As a proof of principle to extract the 1PI vertices from experimentally measured correlations we apply the formalisms discussed above to the physical system of two tunnel-coupled one-dimensional superfluids in a DW potential on an atomchip.
Such a system can be seen as a quantum simulator of the sine-Gordon model~\cite{Gritsev2007,schweigler2017experimental}. The relative phase $\varphi_x$ between the superfluids corresponds to $\hat\Phi$ in Eq.~\eqref{eq:H_SG} in section \ref{sec:SG_thermal} while the relative density fluctuations correspond to the conjugate field $\hat\Pi$.

A schematic of the experimental system is given in Fig.~\ref{fig:Schematics}.
The parameters in $H_\mathrm{SG}$ \eqref{eq:H_SG} are related to the experimental parameters via \mbox{$\lambda_T = 2 \hbar^2 n_\mathrm{1D}/(mk_BT)$}, \mbox{$\ell_J=\sqrt{\hbar/(4mJ)}$}, and \mbox{$g = g_\mathrm{1D} + \hbar J / n_\mathrm{1D}$}. Here the 1D effective interaction strength \mbox{$g_\mathrm{1D} = 2 \hbar a_\mathrm{s} \omega_\perp$} is calculated from the s-wave scattering length $a_\mathrm{s}$ and the frequency $\omega_\perp$ of the radial confinement; $n_\mathrm{1D}$ is the 1D density and $m$ is the mass of the \textsuperscript{87}Rb atoms which the superfluids consist of. The single particle tunneling rate between the wells is denoted by $J$.

In the experiment, the two superfluids are prepared by slow evaporative cooling in the DW potential (the same way the slow cooled data presented in \cite{schweigler2017experimental} was prepared). However, in contrast to \cite{schweigler2017experimental}, the data used here was taken for a box-like longitudinal confinement \cite{Rauereaan7938} of 75{\um} length.
Matter-wave interferometry~\cite{Schumm05} gives access to the spatially resolved relative phase fluctuations $\varphi_x$ between the two superfluids.
More details about the experimental procedure and the data analysis can be found in \cite{schweigler2017experimental,Schweigler19thesis}.

\begin{figure}
	\centering
	\includegraphics{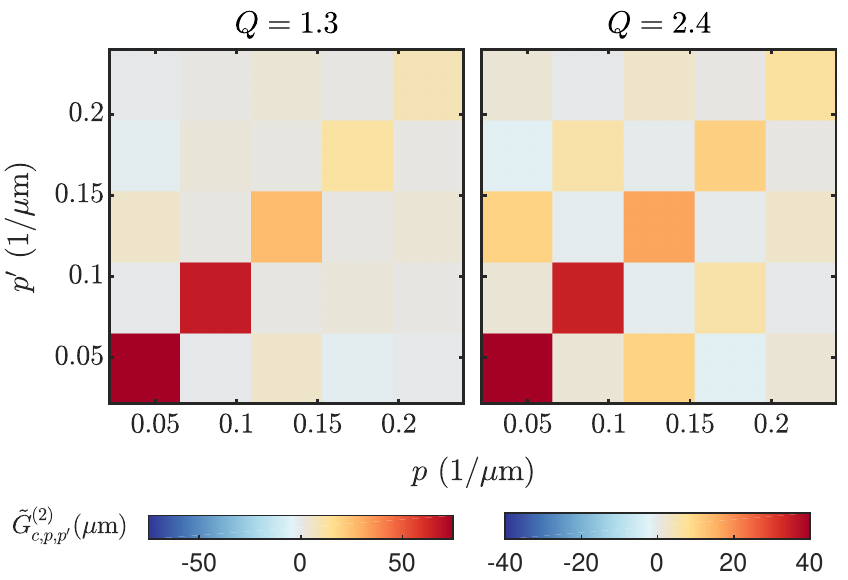}
	\caption{\textbf{Cosine transformed second-order connected correlation function.}
		Results for different phase-locking strength as indicated by the values of $Q$ stated above the respective subplots.
		The color represents the values for cosine transformed second-order connected correlation function $\tilde G^{(2)}_{c,p,p^\prime}$ as defined in \eqref{eq:socond_order_cos_corr}.
		Note that the value (249.9 in the left subplot and 93.9 in the right subplot) for the lowest leftmost data point, lies outside the color-range.
		The color-range was chosen like this to get better visibility.
	}
	\label{fig:2D_2p_cos_corr}
\end{figure}

Starting from the measured phase profiles, we can calculate the 1PI vertices in the same way as was done for the numerics (see section \ref{sec:SG_thermal} and appendix \ref{app:data_analysis}). 
For box like potentials one naturally gets Neumann boundary conditions (BC) for the phase from the condition of vanishing particle current on the edges~\cite{Rauereaan7938}.
From the cosine transform (compatible with the Neumann BC) of the complete system we therefore simply get the 1PI vertices of the Hamiltonian with this BC.
Acknowledging that our system is still too short to get results free from finite size effects, we nevertheless apply the conversion factors to Fourier momentum space given in \eqref{eq:cos_fourier_factor_2p} and \eqref{eq:cos_fourier_factor_4p} for consistency when presenting $\Gamma^{(n)}$.

\begin{figure}
	\centering
	\includegraphics{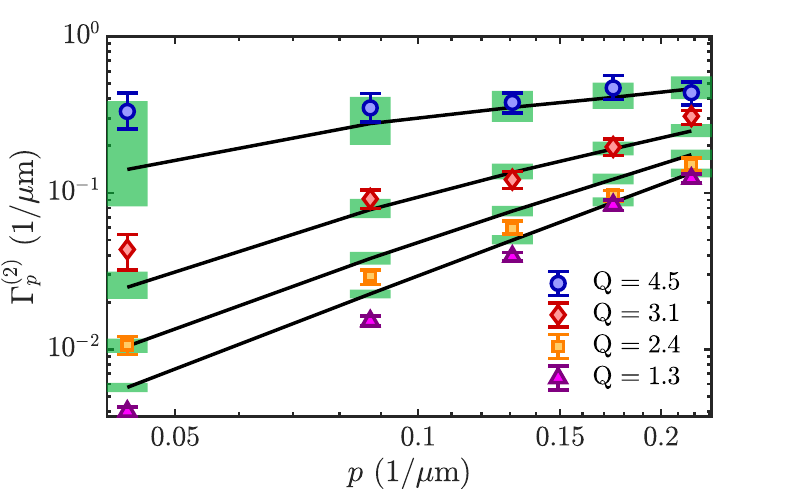}
	\caption{\textbf{Experimental 1PI two-point function.}
		The four different measurements correspond to $Q  = 4.5$ (\bluecircle),  $3.1$ (\reddiamond),  $2.4$ (\orangesquare),  and $1.3$ (\purpletriangle).
		The error bars represent the 80\% confidence intervals obtained using bootstrapping.
		We see good agreement with the theory prediction from the sine-Gordon model in thermal equilibrium calculated for $10^6$ numerical realisations (black solid lines).
		The height of the green bars indicates the 80\% confidence interval for the numerical predictions considering the finite experimental sample size.
		Note that all uncertainty comes from the finite sample size, no uncertainty in the parameters $\lambda_T$ and $Q$ was assumed.
		The width of the bars was chosen arbitrarily.
	}
	\label{fig:propagator_exp}
\end{figure}

Let us start the discussion of the experimental results with the cosine transformed second-order correlation function
\begin{align}
\tilde G^{(2)}_{c,p,p^\prime} = \frac{2}{L} \left( \langle \tilde{\varphi}_p \tilde{\varphi}_{p'} \rangle - \langle \tilde{\varphi}_p \rangle \langle \tilde{\varphi}_{p'} \rangle \right). \label{eq:socond_order_cos_corr}
\end{align} 
Here $\tilde{\varphi}_p$ represents the cosine transform \eqref{eq:cos_trans} over the finite interval with length $L$ and we chose the prefactors for later convenience. The factor 2 comes from the identity \eqref{eq:cos_fourier_factor_2p} and the factor $1/L$ from the delta function.  
We see from Fig.~\ref{fig:2D_2p_cos_corr} that the correlations are approximately diagonal. 
Further, note that density-phase two-point correlations, $\langle \pi \varphi \rangle_{W_t}$, vanish due to time-translation invariance of the thermal state, even beyond the semi-classical approximation Eq.~\eqref{eq:Z_thermal}. Together, this enables us to calculate the 1PI two-point correlator as
\begin{equation}
 \Gamma^{(2)}_p = \frac{1}{\tilde G^{(2)}_{c,p,p}} \; ,
\end{equation}
where we neglected the small off-diagonal elements of $\tilde{G}^{(2)}_{c,p,p^\prime}$.
The results are presented in Fig.~\ref{fig:propagator_exp}.

\begin{figure}
	\centering
	\includegraphics{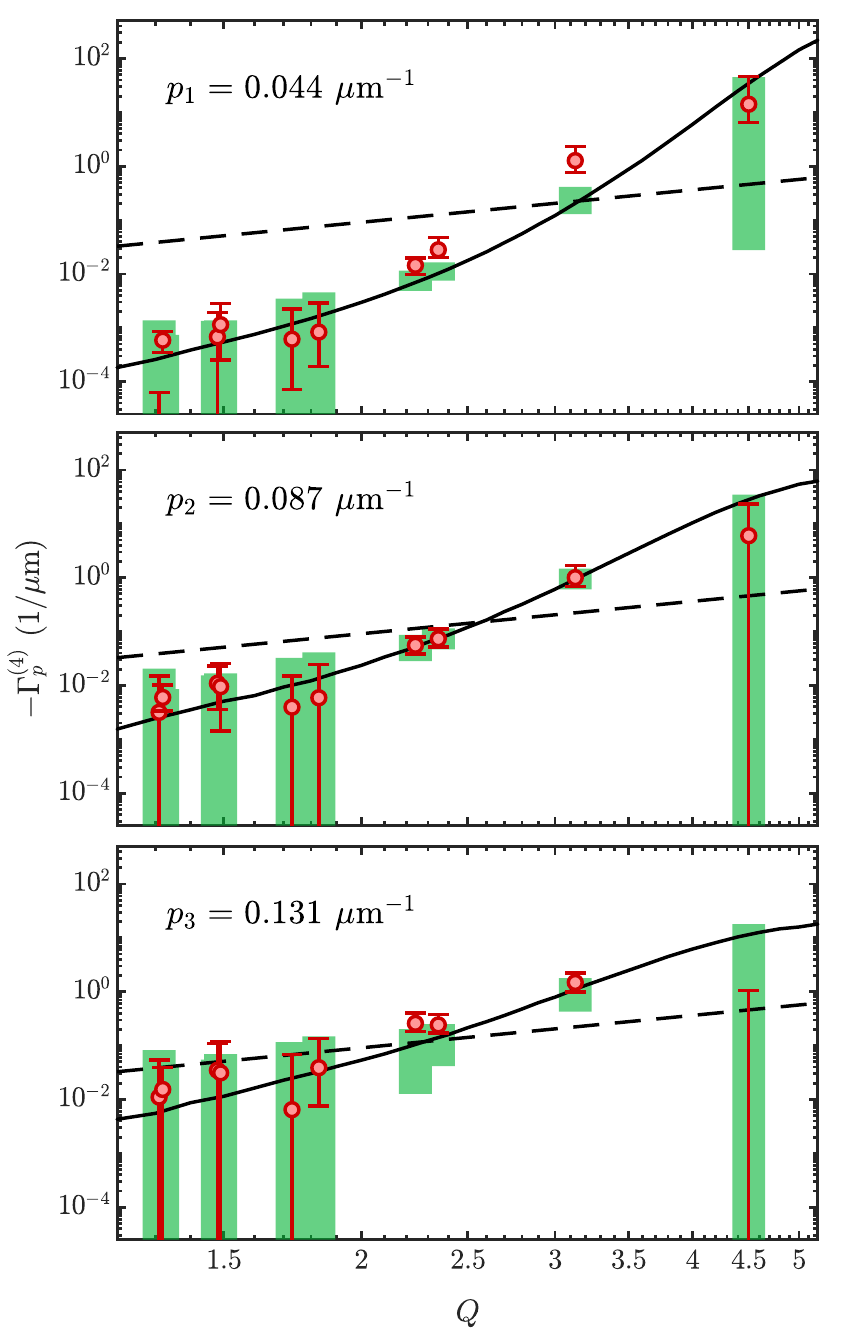}
	\caption{
		\textbf{Experimental 4-Vertex.}
		The red bullets represent the experimental results for the 4-vertex as a function of $Q = \lambda_T/\ell_J$.
		The points in one particular subplot correspond to separate measurements with different tunneling strength.
		The different subplot show the results for the lowest three values of $p$ indicated in the upper left corner of the subplots.
		The error bars represent 80\% confidence intervals obtained using bootstrapping.
		The numerical prediction from the sine-Gordon model in thermal equilibrium is given by the green bars.
		The height of the bars indicates the 80\% confidence interval for the theory predictions considering the finite experimental sample size.
		Note that all uncertainty comes from the finite sample size, no uncertainty in the parameters $\lambda_T$ and $Q$ was assumed.
		The width of the bars was chosen arbitrarily.
		The solid black line represents the theory prediction from $10^6$ numerical realisations and the dashed black line the tree-level prediction~\eqref{eq:tree-level1PI}.
	}
	\label{fig:4p_cos_corr}
\end{figure}

All experimental results presented in this paper are corrected for the expected influence of the finite imaging resolution.
In our simple model, the imaging process leads to a convolution of the true phase profiles with a Gaussian function with $\sigma_\mathrm{psf} = 3${\um} \cite{Schweigler19thesis}.
In momentum space this simply leads to a multiplication with $\exp(-p^2\sigma_\mathrm{psf}^2/2)$, which can be easily corrected by dividing the cosine transformed relative phase $\tilde{\varphi}(p)$ by this factor.

In order to connect the experimental results to the theoretical model (section \ref{sec:SG_thermal}) we estimate $\lambda_T = 11${\um} for all the different measurements. The values for $Q = \lambda_T/\ell_J$ are then self consistently fitted from $\cohfact$~\cite{Schweigler19thesis}. We see good agreement between experiment and thermal sine-Gordon theory for the 1PI two-point function in Fig.~\ref{fig:propagator_exp}.

\begin{figure}[tb]
	\centering
	\includegraphics{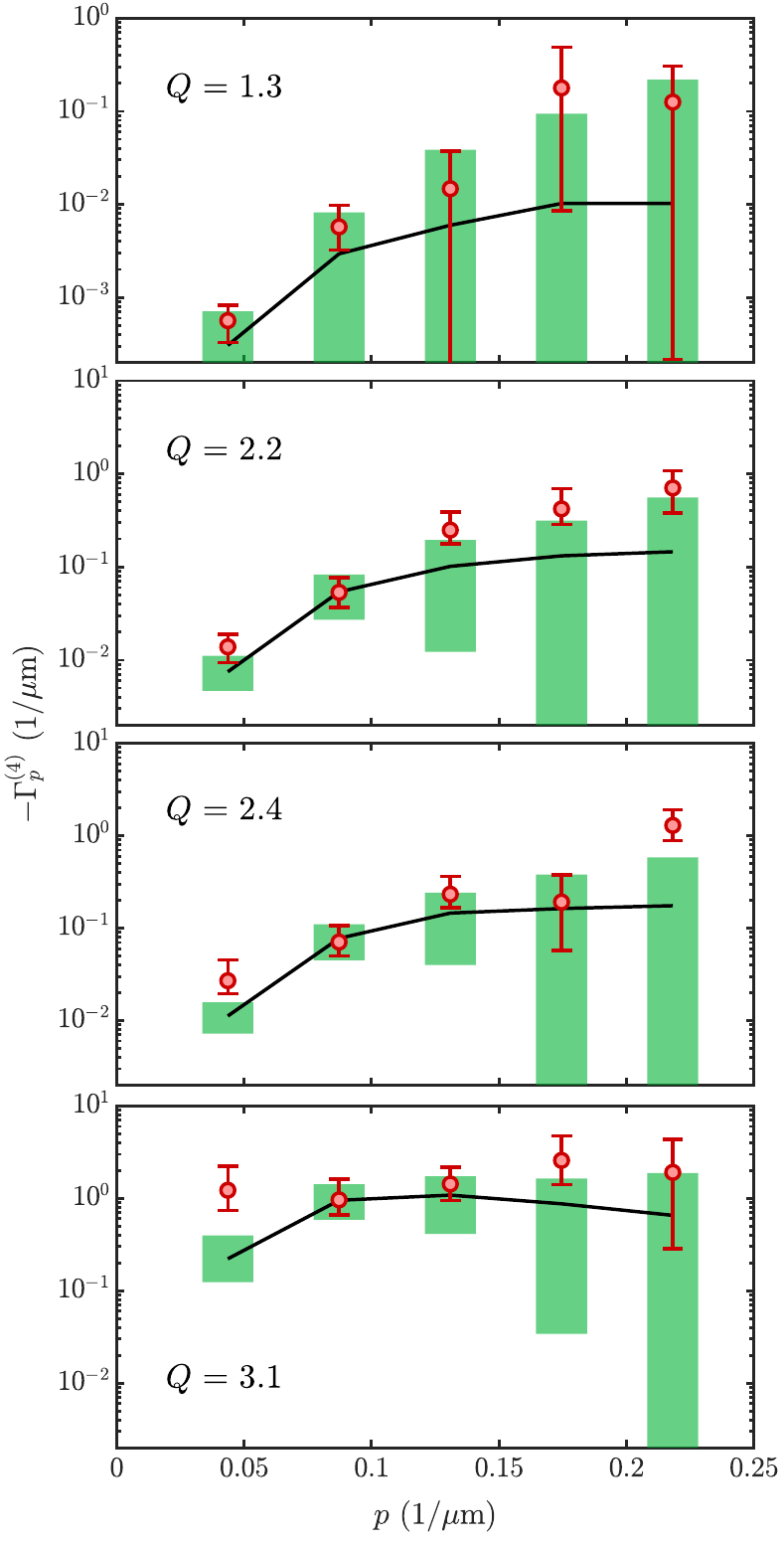}
	\caption{\textbf{Running coupling.}
		Like Fig.~\ref{fig:4p_cos_corr}, but showing $\Gamma^{(4)}_p$ as a function of $p$ for the four measurements with the biggest experimental sample size. 
		Depending on the value of $Q$ indicated in the different subplots one can see a clear momentum dependence, i.e. `running coupling'.
		Note that the vertical axis of the uppermost subplot is different from the rest.
		However, the logarithmic range is the same as in the other subplots. 		
	}
	\label{fig:scan59294pcoscorr}
\end{figure}

Having obtained the two-point function, and using that the third-order correlation functions vanish for symmetry reasons, we can calculated the diagonal part of the 4-vertex as
\begin{equation}
	\Gamma^{(4)}_p = - \frac{8}{3} \frac{1}{L} \langle \tilde{\varphi}_p^4 \rangle_c \times \left(\Gamma^{(2)}_p\right)^4. \label{eq:gamma_4_exp}
\end{equation} 
Here $\langle \tilde{\varphi}_p^4 \rangle_c$ stands for the diagonal elements of the cosine transformed fourth-order connected correlation function.
The factor $8/3$ comes from the identity \eqref{eq:cos_fourier_factor_4p}, the factor $1/L$ again comes from the delta function. 
The results for the three lowest lying momentum modes are presented in Fig.~\ref{fig:4p_cos_corr} as a function of $Q$.
We find qualitative agreement between experiment and theory as well as the expected approach towards the tree-level result for higher momenta.
The momentum dependence for the measurements with large experimental sample size (Fig.~\ref{fig:scan59294pcoscorr}) reveal a running coupling with a qualitative agreement between the experiment and thermal sine-Gordon theory.

\section{\label{sec:conlusion}Conclusion}

The presented method provides a general framework to extract and test the effective or emergent quantum field theoretical description of generic quantum many-body systems from experiments. For the example of the sine-Gordon model, which is quantum simulated with two tunnel-coupled superfluids, we have demonstrated how to experimentally obtain the irreducible vertices in thermal equilibrium and compared to theoretical expectations. This represents an essential step in the verification of the approach, which opens a new pathway to study fundamental questions of QFT through large-scale (analog) quantum simulators.
 
This becomes especially interesting for strongly correlated systems and in non-equilibrium situations, where it is often not possible to solve the theory using classical computational techniques. Extracting the irreducible building blocks of quantum many-body systems, and how they change with time, promises to provide detailed insights into the dynamics for these cases. The recent study of a spin-1 Bose condensate far from equilibrium~\cite{prufer2019experimental}, which has been performed in parallel to this work and employed similar methods, presents an example where currently no theoretical solution is available. In turn, the insight from experimental measurement can support  theoretical developments in devising new approximation schemes and effective field theory descriptions.

\section*{Acknowledgments} 
We thank Thomas Gasenzer, Philipp Kunkel, Igor Mazets, Markus Oberthaler, Maximilian Prüfer, Helmut Strobel and Christof Wetterich for discussions and Federica Cataldini, Sicong Ji, Bernhard Rauer and Mohammadamin Tajik for help with the experiment.
This work is part of and supported by the DFG Collaborative Research Centre ``SFB 1225 (ISOQUANT)'', the ERC Advanced Grant ``QuantumRelax'', and  the JTF project “The Nature of Quantum Networks” (ID 60478). TS acknowledges support by the Austrian Science Fund (FWF) in the framework of the Doctoral School Complex Quantum Systems (CoQuS). SE~acknowledges support through the EPSRC Project Grant (EP/P00637X/1).
JS, JB and SE acknowledge the hospitality of the Erwin Schrödinger Institut in the framework of their thematic program \emph{Quantum Paths} which enabled many discussions defining the foundational ideas shaping this article.

\appendix

\section{Notational conventions}\label{app:notation_summary}

In table \ref{tab:overview}, we have summarized the different generating functionals and the involved fields which appear throughout this paper.
\begin{table}
	\begin{ruledtabular}
		\begin{tabular}{ll}
			$\hat{\rho}[\hat{\Phi}]$ & density operator\\
			$W[\varphi]$ & Wigner functional \\
			$Z[J]$ & g.f. for full correlators $G^{(n)}$ \\
			$E[J]$ & g.f. for connected correlators $G_c^{(n)}$ \\
			$\Gamma[\Phi]$ & g.f. for 1PI correlators $\Gamma^{(n)}$ \\
			&\\
			$\hat{\Phi}$ & field operator (non-commuting) \\
			$\varphi$ & fluctuating (microscopic) field \\
			$J$ & auxiliary source field \\
			$\Phi$ & macroscopic field
		\end{tabular}
	\end{ruledtabular}
	\caption{\label{tab:overview}Overview over the different objects that are appear in the general discussion. `g.f.' abbreviates `generating functional'.}
\end{table}

We use the following notations: Operators are always indicated by a hat. $\text{Tr}\left[\dots \right]$ indicates a trace over the full Hilbert space. The absence of a hat implies a c-number (i.e.~commuting objects). In the whole formalism, the time $t$ is treated as a label and often left out for brevity. Repeated spatial indices are integrated over, e.g.~\mbox{$J^\varphi_\mathbf{x} \hat{\Phi}_\mathbf{x} = \int d^dx \, J^\varphi(\mathbf{x}) \hat{\Phi}(\mathbf{x})$}; we write explicit integrals if there is room for confusion. 

In the appendix, we also use collective latin indices, $a = (\varphi/\pi, \mathbf{x})$; then repeated indices are integrated or summed over as appropriate, e.g.~$J_a \hat{\Phi}_a = \int d^dx \left[J^\varphi(\mathbf{x}) \hat{\Phi}(\mathbf{x}) + J^\pi(\mathbf{x}) \hat{\Pi}(\mathbf{x})\right]$. It is useful to think of correlation functions as tensors, e.g.~$G^{(2)}_{c,ab}$ and its inverse $\Gamma^{(2)}_{ab}$ are the components of 2-tensors that fulfill $G^{(2)}_{c,ab} \Gamma^{(2)}_{bd} = \delta_{ad}$. Here $\delta_{ad}$ is the product of a discrete Kronecker delta and a continuous Dirac delta distribution.

\section{Technical details of section \ref{sec:ET_QFT}}
In this appendix we show explicit calculations and detailed dicussions that we left out in the discussion of equal-time correlation functions.

\subsection{Operator ordering at equal-time\label{app:ordering}}
There are three obvious choices of different orderings,
often referred to as symmetric (Weyl), normal (P) and anti-normal (Q). In terms of creation and annihiliation operators $\hat{a}_p$ and $\hat{a}^\dagger_p$, which fulfill $\left[\hat{a}_p, \hat{a}^\dagger_q\right] \sim \delta_{pq}$, they can be realized by the definitions
\begin{subequations}
	\begin{align}
	Z^{(W)}_t[J] &=  \text{Tr} \left[ \hat{\rho}_t \exp \left(J_p \hat{a}^\dagger_p - J^*_p \hat{a}_p \right)\right] \; ,\\
	Z^{(P)}_t[J] &=  \text{Tr} \left[ \hat{\rho}_t \exp \left(J_p \hat{a}^\dagger_p \right) \exp\left(  - J^*_p \hat{a}_p \right)\right] \; ,\\
	Z^{(Q)}_t[J] &=  \text{Tr} \left[ \hat{\rho}_t  \exp\left(  - J^*_p \hat{a}_p \right) \exp \left(J_p \hat{a}^\dagger_p \right) \right] \; .
	\end{align}
\end{subequations}
In general, there is a continuum of other choices that smoothly connect these three cases.
However, all different choices are fully equivalent in the sense that they contain all measurable information and the main difference lies in the associated quasi-probability distributions. For more details we refer to~\cite{cahill1969density}.

Explicitly, for the choice of the main text, the ordering is resolved as
\begin{subequations}
\begin{align}
\exp \left(J^\varphi_\mathbf{x} \hat{\Phi}_\mathbf{x} + J^\pi_\mathbf{x} \hat{\Pi}_\mathbf{x} \right) &= e^{J^\varphi_\mathbf{x} \hat{\Phi}_\mathbf{x}} e^{J^\pi_\mathbf{x} \hat{\Pi}_\mathbf{x}}  e^{-\frac{i}{2} J^\varphi_\mathbf{x}J^\pi_\mathbf{x}}\\
&= e^{J^\pi_\mathbf{x} \hat{\Pi}_\mathbf{x}}e^{J^\varphi_\mathbf{x} \hat{\Phi}_\mathbf{x}}   e^{\frac{i}{2} J^\varphi_\mathbf{x}J^\pi_\mathbf{x}}
\end{align}
\end{subequations}
where we used the BCH formula in the form
\begin{align}
e^{\hat{A}+\hat{B}} = e^{\hat{A}} e^{\hat{B}} e^{-\frac{1}{2}\left[\hat{A},\hat{B}\right]} \; ,
\end{align}
which is valid for $\left[\left[\hat{A},\hat{B}\right],\hat{B}\right]=\left[\left[\hat{A},\hat{B}\right],\hat{A}\right]= 0$. Thus, derivatives acting on $\exp \left(J^\varphi_\mathbf{x} \hat{\Phi}_\mathbf{x} + J^\pi_\mathbf{x} \hat{\Pi}_\mathbf{x} \right)$ from the left result in operators and additional sources according to
\begin{align}
	\frac{\delta}{\delta J^\varphi_\mathbf{x}} \rightarrow \hat{\Phi}_\mathbf{x} - \frac{iJ^\pi_\mathbf{x}}{2} \;, && 	\frac{\delta}{\delta J^\pi_\mathbf{x}} \rightarrow \hat{\Pi}_\mathbf{x} + \frac{iJ^\varphi_\mathbf{x}}{2} \; .
\end{align}
Using this correspondence, it is straightforward to generate explicit expression for all correlators. For instance, at second order, we have
\begin{align}
	\frac{\delta}{\delta J^\varphi_\mathbf{x}} \frac{\delta}{\delta J^\pi_\mathbf{y}} &\rightarrow \frac{\delta}{\delta J^\varphi_\mathbf{x}} \left(  \hat{\Pi}_\mathbf{y} + \frac{iJ^\varphi_\mathbf{y}}{2} \right) \\
	&\rightarrow \left(  \hat{\Pi}_\mathbf{y} + \frac{iJ^\varphi_\mathbf{y}}{2} \right)  \left(\hat{\Phi}_\mathbf{x} - \frac{iJ^\pi_\mathbf{x}}{2}\right)+ \frac{i}{2} \delta \left(\mathbf{x}- \mathbf{y}\right) \; . \nonumber
\end{align}
Setting the sources to zero proves that
\begin{align}
	\left\langle\varphi_{\mathbf{x}} \pi_{\mathbf{y}} \right\rangle_{W_t} &= \text{Tr} \left[ \hat{\rho}_t \hat{\Pi}_\mathbf{y} \hat{\Phi}_\mathbf{x} \right] + \frac{i}{2} \delta \left(\mathbf{x}- \mathbf{y}\right) \nonumber \\ &=  \frac{1}{2}\text{Tr} \left[ \hat{\rho}_t \left(\hat{\Phi}_\mathbf{x} \hat{\Pi}_\mathbf{y} + \hat{\Pi}_\mathbf{y} \hat{\Phi}_\mathbf{x} \right)  \right] \;,
\end{align}
where we used the canonical commutation relations and the normalization of $\hat{\rho}_t$.
In the following we drop the label $t$ for brevity.

\subsection{\label{app:rho_from_correlations}Correlations and the density operator}
The density operator $\hat{\rho}$ can formally be recovered from $Z$ as~\cite{cahill1969density}
\begin{align}
	\hat{\rho} &= \int \mathcal{D} J^\varphi  \mathcal{D} J^\pi \, Z[J]\left[\exp \left(J^\varphi_\mathbf{x} \hat{\Phi}_\mathbf{x} + J^\pi_\mathbf{x} \hat{\Pi}_\mathbf{x} \right) \right]^{-1} \; .
\end{align}
Furthermore, the mappings between the different functionals are invertible (under appropriate mathematical assumptions): $W$ and $Z$ are related by Fourier transforms, $Z$ and $E$ by an exponential (or logarithmic) map, $E$ and $\Gamma$ by Legendre transforms.
Thus, it is completely equivalent to work with the Wigner functional $W$ or any of the generating functionals $Z$, $E$, $\Gamma$ instead of the density operator $\hat{\rho}$.

\subsection{\label{app:Zt_with_Wt}Functional integral representation, Eq.~\eqref{eq:Zt_with_Wt}}

We seek a representation of $Z$ in terms of classical (commuting) instead of operator-valued fields. To this end, we evaluate the trace as
\begin{widetext}
\vspace*{-4mm}
	\begin{subequations}
		\begin{align}
		Z[J,\rho(t)] 
		&= \int \mathcal{D}\varphi^+ \mathcal{D}\varphi^-  
		\matrixelement{\varphi^+}{\rho(t)}{\varphi^-}
		\matrixelement{\varphi^-}{e^{J^\varphi_\mathbf{x} \Phi_\mathbf{x} + J^\pi_\mathbf{x} \Pi_\mathbf{x}}}{\varphi^+} \\
		&= \int \mathcal{D}\varphi^+ \mathcal{D}\varphi^-\, \mathcal{D}\tilde{\pi}\, 
		\matrixelement{\varphi^+}{\rho(t)}{\varphi^-}
		\matrixelement{\varphi^-}{e^{J^\varphi_\mathbf{x} \Phi_\mathbf{x}}}{\tilde{\pi}}
		\matrixelement{\tilde{\pi}}{e^{J^\pi_\mathbf{x} \Pi_\mathbf{x}}}{\varphi^+}e^{-\frac{i}{2}J^\varphi_\mathbf{x}J^\pi_\mathbf{x}} \\
		&= \int \mathcal{D}\varphi \, \mathcal{D}\tilde{\varphi}\, \mathcal{D}\tilde{\pi}\,
		\matrixelement{\varphi + \frac{\tilde{\varphi}}{2}}{\rho(t)}{\varphi - \frac{\tilde{\varphi}}{2}}
		e^{J^\varphi_\mathbf{x} \left(\varphi_\mathbf{x}-\frac{\tilde{\varphi}_\mathbf{x}}{2}\right)+i \left(\varphi_\mathbf{x}-\frac{\tilde{\varphi}_\mathbf{x}}{2}\right)\tilde{\pi}_\mathbf{x}+J^\pi_\mathbf{x} \tilde{\pi}_\mathbf{x}-i \left(\varphi_\mathbf{x}+\frac{\tilde{\varphi}_\mathbf{x}}{2}\right)\tilde{\pi}_\mathbf{x}-\frac{i}{2}J^\varphi_\mathbf{x}J^\pi_\mathbf{x}}\\
		&= \int \mathcal{D}\varphi \, \mathcal{D}\tilde{\varphi}\, \mathcal{D}\tilde{\pi}\, \mathcal{D}\pi\,
		W_t\left[\varphi,\pi\right]
		e^{i\pi_\mathbf{x}\tilde{\varphi}_\mathbf{x}+J^\varphi_\mathbf{x} \left(\varphi_\mathbf{x}-\frac{\tilde{\varphi}_\mathbf{x}}{2}\right)+J^\pi_\mathbf{x} \tilde{\pi}_\mathbf{x}-i\tilde{\varphi}_\mathbf{x}\tilde{\pi}_\mathbf{x}-\frac{i}{2}J^\varphi_\mathbf{x}J^\pi_\mathbf{x}}\\
		&= \int \mathcal{D}\varphi \, \mathcal{D}\pi\, 
		W_t\left[\varphi,\pi\right] \exp \left[J^\varphi_\mathbf{x} \varphi_\mathbf{x}+J^\pi_\mathbf{x} \pi_\mathbf{x}\right] \; .
		\end{align}
	\end{subequations}
\vspace*{-4mm}
\end{widetext}
In the above calculation, we have again used the BCH formula, performed a change of variables $\varphi^\pm \equiv \varphi \pm \tilde{\varphi}/2$, and employed the definition of the Wigner functional.

\subsection{\label{app:conn_from_full}The connected correlators, Eq.~\eqref{eq:howto_connected}}
To discuss the explicit form of the connected and 1PI correlators, we use the short-hand notation, where sources $J_a$ have a single index indicating space $\mathbf{x}$, as well as any of the two fields $\varphi$ and $\pi$. Repeated indices are summed and integrated over. Similarly, we abbreviate the fields as $\varphi_a$.

At first order, the connected correlators are directly related to the full Weyl-ordered one-point function,
\begin{align}\label{eq:E1}
	\frac{\delta E}{\delta J_a} = \frac{\delta \log Z}{\delta J_a} = \frac{1}{Z}	\frac{\delta Z}{\delta J_a}\; .
\end{align}
Setting the sources to zero, we have $Z[J=0]=1$, which proves Eq.~\eqref{eq:connected_a}.

At second order, we calculate
\begin{align}\label{eq:E2}
\frac{\delta^2 E}{\delta J_a\delta J_b} = \frac{\delta}{\delta J_a} \left(\frac{1}{Z}	\frac{\delta Z}{\delta J_b}\right) = \frac{1}{Z}	\frac{\delta^2 Z}{\delta J_a \delta J_b}  -  \frac{1}{Z^2}	\frac{\delta Z}{\delta J_a}	\frac{\delta Z}{\delta J_b}\; ,
\end{align}
which proves Eq.~\eqref{eq:connected_b}.

The higher orders follow analogously by the combinatorics of the derivatives. E.g., the third order, Eq.~\eqref{eq:connected_c}, is obtained by
\begin{align}
	\frac{\delta^3 E}{\delta J_a\delta J_b \delta J_c} &= \frac{1}{Z}	\frac{\delta^3 Z}{\delta J_a \delta J_b \delta J_c} \nonumber\\ &- \left(\frac{1}{Z^2}	\frac{\delta^2 Z}{\delta J_a \delta J_b} \frac{\delta Z}{\delta J_c} + \text{2 perm.}\right) \nonumber \\
	&+2  \frac{1}{Z^3}	\frac{\delta Z}{\delta J_a}	\frac{\delta Z}{\delta J_b}\frac{\delta Z}{\delta J_c} \; 
\end{align}
and using Eqs.~\eqref{eq:E1} and \eqref{eq:E2}.

\subsection{\label{app:1PI_from_conn}The 1PI vertices, Eq.~\eqref{eq:explicit1PI}}
The expression for the 1PI two-point function is central for the construction of the higher orders. It follows by considering a derivative of the stationarity condition,
\begin{align}
	\frac{\delta^2 \Gamma}{\delta \Phi_a \delta \Phi_b} = \frac{\delta J_b}{\delta \Phi_a}  \;.
\end{align}
This is the matrix inverse of the derivative of the one-point function (in the presence of sources)
\begin{align}
	\frac{\delta \Phi_a(J)}{\delta J_b} =\frac{\delta}{\delta J_b} \left(\frac{1}{Z}	\frac{\delta Z}{\delta J_a}\right) =  \frac{\delta^2 E}{\delta J_a \delta J_b} \; .
\end{align}
Thus, we find Eq.~\eqref{eq:1PI_b}, or
\begin{align}
	\frac{\delta^2 \Gamma}{\delta \Phi_a\delta \Phi_b} = \left[\left(\frac{\delta^2 E}{\delta J \delta J}\right)^{-1}\right]_{ab} \; ,
\end{align}
which also holds without setting the sources to zero.

The higher orders follow by taking derivatives of this equation (with non-zero sources). To this end, we replace a derivative by
\begin{align}
\frac{\delta}{\delta \Phi_c} = 	\frac{\delta J_{c'}}{\delta \Phi_c} \frac{\delta}{\delta J_{c'}} = 	\frac{\delta^2 \Gamma}{\delta \Phi_c \delta \Phi_{c'}}  \frac{\delta}{\delta J_{c'}} 
\end{align}
and calculate the derivative of the inverse of a matrix $M(y)$ depending on a parameter $y$ according to
\begin{align}
\frac{d}{dy} \left( M^{-1} \right) = - M^{-1} \cdot \frac{dM}{dy} \cdot M^{-1} \; . 
\end{align}
This results for the third order in
\begin{align}
		\frac{\delta^3 \Gamma}{\delta \Phi_a\delta \Phi_b \delta \Phi_c} = - \frac{\delta^2 \Gamma}{\delta \Phi_a \delta \Phi_{a'}}  \frac{\delta^2 \Gamma}{\delta \Phi_b \delta \Phi_{b'}}  \frac{\delta^2 \Gamma}{\delta \Phi_c \delta \Phi_{c'}} \frac{\delta^3 E}{\delta J_{a'} \delta J_{b'} \delta J_{c'}} \;,
\end{align}
which proves Eq.~\eqref{eq:1PI_c}.

Similarly, the fourth order, Eq.~\eqref{eq:1PI_d} , follows by the combinatorics of taking further derivatives,
\begin{widetext}
	\begin{align}
	\frac{\delta^4 \Gamma}{\delta \Phi_a\delta \Phi_b \delta \Phi_c \delta \Phi_d} = &- \frac{\delta^2 \Gamma}{\delta \Phi_a \delta \Phi_{a'}}  \frac{\delta^2 \Gamma}{\delta \Phi_b \delta \Phi_{b'}}  \frac{\delta^2 \Gamma}{\delta \Phi_c \delta \Phi_{c'}}   \frac{\delta^2 \Gamma}{\delta \Phi_d \delta \Phi_{d'}} \frac{\delta^4 E}{\delta J_{a'} \delta J_{b'} \delta J_{c'}  \delta J_{d'}} \nonumber\\
	 &+ \frac{\delta^2 \Gamma}{\delta \Phi_a \delta \Phi_{a'}}  \frac{\delta^2 \Gamma}{\delta \Phi_b \delta \Phi_{b'}}  \frac{\delta^2 \Gamma}{\delta \Phi_c \delta \Phi_{c'}}   \frac{\delta^2 \Gamma}{\delta \Phi_d \delta \Phi_{d'}}  \left( \frac{\delta^3 E}{\delta J_{a'} \delta J_{b'} \delta J_{e}}  \frac{\delta^2 \Gamma}{\delta \Phi_e \delta \Phi_{f}}  \frac{\delta^3 E}{\delta J_{f} \delta J_{c'} \delta J_{d'}}  + \text{2 perm.}\right) \; .
	\end{align}
\end{widetext}

\subsection{\label{app:W_thermal}The thermal case and the classical limit}
In thermal equilibrium, the (unnormalized) canonical density operator $\hat{\rho}_\beta = e^{-\beta \hat{H}}$ fulfills the equation \mbox{$\partial_\beta \hat{\rho}_\beta = -\frac{1}{2} \left(\hat{H} \hat{\rho}_\beta + \hat{\rho}_\beta  \hat{H} \right)$}. Employing the quasi-probability formalism~\cite{hillery1984distribution}, one can show that this equation translates to an equation for $W_\beta$. It takes the form
\begin{align}
	\partial_\beta W_\beta = -\frac{1}{2} \left[ H_W^+ + H_W^- \right] W_\beta \;,
\end{align}
where $H^\pm_W = H_W\left[\varphi \pm \frac{i\hbar}{2} \frac{\delta}{\delta \pi}, \pi \mp \frac{i\hbar}{2} \frac{\delta}{\delta \varphi}\right]$ is a functional differential operator obtained from the Weyl-transform $H_W[\varphi, \pi]$ of the Hamiltonian $\hat{H}$ by replacing the arguments with the given operators~\cite{hillery1984distribution}. With the initial conditon $W_{\beta \rightarrow \infty} = \text{const.}$, which follows from the high-temperature limit, this functional flow equation can be solved perturbatively by exanding \mbox{$W_\beta = \exp \left[ \sum_{n=0}^{\infty} \hbar^n W_\beta^{(n)} \right]$} in powers of $\hbar$ and comparing the coefficients. 

The first order in this expansion is the classical field theory limit, where $W_\beta \sim e^{-\beta H}$ with the classical Hamiltonian $H = H_W$. Parametrically, this is a valid approximation when $\hbar$ is small.
Since $\hbar$ is a dimensionfull quantity, the precise power-counting of this expansion has to be determined for each theory separately. In general, any quantum system in thermal equilibrium will be governed by (at least) two dimensionless parameters $\epsilon_\text{q}$ and $\epsilon_\text{th}$ that control the strength of quantum and classical fluctuations, respectively. A sufficient condition for the validtiy of the classical approximation is
\begin{align}
	\epsilon_\text{q} \ll \operatorname{min}[1,\epsilon_\text{th}] \; .
\end{align}

The parameters $\epsilon_\text{q}$ and $\epsilon_\text{th}$ are obtained by rescaling the Hamiltonian and the fundamental fields to dimensionless quantities. Explicitly, for the sine-Gordon model in the form of Eq.~\eqref{eq:H_SG}, we have
\begin{align}
\beta\hat{H} &= \frac{1}{\epsilon_\text{th}} \int dx' \left\lbrace \frac{1}{2} \left[\left(\hat{\Pi}'\right)_{x'}^2 + \left(\partial_{x'} \hat{\Phi}'_{x'}\right)^2 \right] - \cos \left(\hat{\Phi}'_{x'}\right)\right\rbrace \;,
\end{align}
together with the rescaled commutation relations
\begin{align}
	\left[\hat{\Phi}'_{x'}, \hat{\Pi}'_{y'}\right] &= i \epsilon_\text{q} \delta(x'-y') \; ,
\end{align}
such that we find
\begin{align}
	\epsilon_\text{q} =  \sqrt{4 \gamma} \;, && \epsilon_{\text{th}} = \frac{4 \ell_J}{\lambda_T} = \frac{4}{Q} \; .
\end{align}
Here
\begin{align}
	\gamma = \frac{mg}{\hbar^2 n_\text{1D}} = \gamma_\mathrm{LL} + \frac{1}{n_\text{1D}^2 \ell_J^2} \; ,
\end{align}
is dominated by the 1D Lieb-Liniger parameter \mbox{$\gamma_\mathrm{LL} = mg_\mathrm{1D}/(\hbar^2 n_\text{1D})$}, such that the semi-classical approximation is valid in the weakly interacting regime $\gamma_\text{LL} \ll 1$, as expected.

\subsection{\label{app:separation_phi_and_pi}Derivation of Eq.~\eqref{eq:separation_of_phi_and_pi} and the $\Pi$-dependence} 
From Eq.~\eqref{eq:Z_thermal}, the generating functional separates into a product $Z_\beta[J] = Z^\varphi_\beta[J^\varphi] Z^\pi_\beta[J^\pi]$ with
\begin{subequations}
	\begin{align}
	Z^\varphi_\beta[J^\varphi] &\sim \int \mathcal{D}\varphi\, e^{-\beta \int_\mathbf{x} \left[\frac{1}{2} \left(\nabla_\mathbf{x}\varphi_\mathbf{x}\right)^2 + V_\mathbf{x}(\varphi)\right] + \int_\mathbf{x} J^\varphi_\mathbf{x}\varphi_\mathbf{x}} \;,\\
	Z^\pi_\beta[J^\pi] &\sim  \int \mathcal{D}\pi\, \exp \left[-\frac{\beta}{2} \int_\mathbf{x}  \pi_\mathbf{x}^2  + \int_\mathbf{x} J^\pi_\mathbf{x}\pi_\mathbf{x}\right]  \nonumber\\ &\sim  \exp \left[\frac{1}{2\beta} \int_{\mathbf{x}} \left(J^\pi_\mathbf{x}\right)^2 \right]  \; .
	\end{align}
\end{subequations}
This directly implies that $E[J] = \log Z^\varphi_\beta[J^\varphi] + \log Z^\pi_\beta[J^\pi] + \text{const.}$ and thus the effective action becomes $\Gamma[\Phi,\Pi] = \Gamma^\varphi[\Phi] + \Gamma^\pi[\Pi] + \text{const.}$, which proves Eq.~\eqref{eq:separation_of_phi_and_pi}. 

Carrying out the Legendre transform in $J^\pi$, we solve
\begin{align}
	\Pi_\mathbf{x}(J^\pi) = \left.\frac{\delta Z^\pi[J^\pi]}{\delta J_\mathbf{x}^\pi}\right|_{J=0} = \frac{J_\mathbf{x}^\pi}{\beta} \Rightarrow J_\mathbf{x}^\pi(\Pi) = \beta \Pi_\mathbf{x}\;.
\end{align}
and finally obtain
\begin{align}
	\Gamma^\pi[\Pi] = - \log Z^\pi[J^\pi(\Pi)] + J_\mathbf{x}^\pi(\Pi) \Pi _\mathbf{x}= \frac{\beta}{2} \int_\mathbf{x} \Pi_\mathbf{x}^2 \; .
\end{align}

\subsection{Time translation invariance}\label{app:B8}
For a stationary system, all observables are time-independent, $\partial_t \text{Tr} \left[ \hat{\rho}_t \dots \right] = 0$. If additionally the Hamiltonian is of the form $\hat{H}[\hat{\Phi}, \hat{\Pi}] = \hat{H}[\hat{\Pi}]+ \hat{H}[\hat{\Phi}]$ with $\hat{H}[\hat{\Pi}] = \frac{1}{2}\int_{\mathbf{x}}  \Pi_{\mathbf{x}}^2$, then it follows that $	0 = \text{Tr} \left[ \hat{\rho}_t \left( \hat{\Phi}_x \hat{\Pi}_y + \hat{\Pi}_x \hat{\Phi}_y\right) \right]$
and further $\langle \varphi_x \pi_y \rangle_{W_t} = 0$. As a consequence the two-point function becomes block diagonal, which simplifies the inversion,
\begin{align} \label{eq:inverse_2pt}
	G^{(2)}_c = \begin{pmatrix}
	\langle \varphi \varphi \rangle & 0 \\
	0 & \langle \pi \pi \rangle
	\end{pmatrix} \Rightarrow 
	\Gamma^{(2)} = \begin{pmatrix}
	\langle \varphi \varphi \rangle^{-1} & 0 \\
	0 & \langle \pi \pi \rangle^{-1}
	\end{pmatrix} \; .
\end{align}

\subsection{\label{app:derivation_loop_expansion}Loop expansion of the effective action}
Starting from Eq.~\eqref{eq:measure_H}, we calculate
\begin{subequations}
	\begin{align}
	e^{-\Gamma_\beta'[\Phi]} &= e^{-\Gamma_\beta[\Phi] + \beta H[\Phi]} = e^{\log Z[J(\Phi)]-J_\mathbf{x}(\Phi) \Phi_\mathbf{x} + \beta H[\Phi]} \\
	&= \int \mathcal{D} \varphi \, e^{-\beta H[\varphi] + J_\mathbf{x}(\Phi) \varphi_\mathbf{x}-J_\mathbf{x}(\Phi) \Phi_\mathbf{x} + \beta H[\Phi]}\\
	&= \int \mathcal{D} \varphi \, e^{-\beta \left(H[\varphi+\Phi] - H[\Phi]\right) +J_\mathbf{x}(\Phi) \varphi_\mathbf{x}  }\\
	&= \int \mathcal{D} \varphi \, e^{-\beta \left(H[\varphi+\Phi] - H[\Phi] - \frac{\delta H[\Phi]}{\delta \Phi_\mathbf{x}} \varphi_\mathbf{x} \right) +\frac{\delta \Gamma_\beta'[\Phi]}{\delta \Phi_\mathbf{x}} \varphi_\mathbf{x}  } \;.
	\end{align}
\end{subequations}
Here, we have used Eqs.~\eqref{eq:eff_action} and \eqref{eq:Z_thermal}, then performed a changed of variables $\varphi \rightarrow \varphi + \Phi$ and finally expressed the sources as $J_\mathbf{x}(\Phi) = \left(\delta \Gamma[\Phi] \right) / \left(\delta \Phi_\mathbf{x}\right)$ .

The first non-trivial correction (one-loop) is obtained by neglecting all terms beyond quadratic order in the fluctuating fields $\varphi$. The remaining gaussian integral can then be performed analytically, which gives
\begin{align}
	e^{-\Gamma'^{\text{,one-loop}}_\beta[\Phi]} &= \int \mathcal{D} \varphi \, e^{-\frac{1}{2} \varphi_\mathbf{x}  G^{-1}_{\mathbf{x},\mathbf{y}}[\Phi] \varphi_\mathbf{y}} \nonumber \\
	&= \left(\det G^{-1}[\Phi] \right)^{-1/2} \;.
\end{align}
The name `one-loop' stems from the expansion in terms of the tree-level two-point function $G_0$. To see this, we rewrite
\begin{subequations}\label{eq:explicit_one-loop}
	\begin{align}
	\Gamma'^{\text{,one-loop}}_\beta[\Phi] &=  \frac{1}{2}\log \det  G^{-1}[\Phi]  = \frac{1}{2}\text{Tr} \log  G^{-1}[\Phi]  \\
	&= \frac{1}{2}\text{Tr} \log \left(G_0 G^{-1}[\Phi]  \right) + \frac{1}{2}\text{Tr} \log   G_0^{-1}\\
	&= -\frac{1}{2} \sum_{n=1}^{\infty} \frac{(-1)^n}{n} \text{Tr}  \left\lbrace G_0 \left( G^{-1}[\Phi] - G_0^{-1} \right)\right\rbrace^n
	\end{align}
\end{subequations}
where we have used the identity $\log \det A = \text{Tr} \log A$, employed the series expansion of the logarithm and dropped the irrelevant constant. Graphically the result can be pictured as a sum of loops consisting of lines that stand for $G_0$ connected by field insertions coming from $ \left( G^{-1}[\Phi] - G^{-1}_0 \right)$.
For more details about the loop expansion, we refer to~\cite{weinberg1995quantum}.

Note that in the standard (unequal-time) formalism, the loop expansion is used as an expansion in weak quantum fluctuations. Here, in the context of the classical field theory limit in thermal equilibrium, it is employed as an expansion in weak thermal fluctuations. In terms of the dimensionless parameters introduced in section~\ref{app:W_thermal}, the loop expansion is applicable for $\epsilon_{\text{th}} \ll 1$. Thus, the tree-level approximation (i.e.~the leading order in the loop expansion), which is used in the main text to extract the microscopic Hamiltonian parameters, is applicable when the following separation of scales holds:
\begin{align}
	\epsilon_{q} \ll \epsilon_{\text{th}} \ll 1 \; .
\end{align}
Colloquially speaking, this is the limit of weak thermal flucuations and even weaker quantum fluctuations.

\section{Details about the data analysis}
\label{app:data_analysis}

This section contains more technical details concerning the practical example, which is discussed in the main text.

\subsection{Cosine vs. Fourier transform and the boundary conditions}
For an infinite system with translation invariance, the correlation functions in (Fourier) momentum space are directly related to the correlators obtained by a cosine transform. Explicitly, with the transforms
\begin{subequations}
	\begin{align}
	\varphi_p &= \int dx \, e^{-ipx} \varphi_x \;, \\
	\tilde{\varphi}_p &= \int dx \, \cos(px)\varphi_x = \frac{1}{2} \left(\varphi_p + \varphi_{-p}\right) \;, \label{eq:cos_trans}
	\end{align}
\end{subequations}
the two-point functions are related as
\begin{align}
\langle \tilde{\varphi}_p^2 \rangle = \frac{1}{2}\langle \varphi_p \varphi_{-p} \rangle \;, \label{eq:cos_fourier_factor_2p}
\end{align}
where we assumed translation invariance, thus $\langle \varphi_p^2 \rangle = 0$.
Similarly, for the four-point functions, we have
\begin{align}
\langle \tilde{\varphi}_p^4 \rangle = \frac{3}{8} \langle \varphi_p \varphi_p \varphi_{-p} \varphi_{-p}  \rangle \; , \label{eq:cos_fourier_factor_4p}
\end{align}
where the prefactors arises from the $6$ nonvanishing contributions out of $2^4 = 16$ combinations. 

In practice, we deal with a finite system without periodic boundary conditions. A discrete Fourier transform is then not appropriate as it yields numerical artifacts. Since the calculation of 1PI correlators simplifies tremendously in Fourier space, we still prefer to work in a Fourier basis. Therefore, we calculate the correlators with a discrete cosine transform, which reduced the artifacts from the boundary conditions. Then we translate the correlators using the factors of $1/2$ and $3/8$ to Fourier-space correlators and subsequently calculate the 1PI correlation functions. For sufficiently large system sizes, this procedure yields the desired results and reduces numerical artifacts in a controlled way.

\subsection{The 1PI vertices from the numerical data}
In practice, the numerical and experimental profiles live on a spatial lattice with lattice spacing $\Delta x$ and a finite number of lattice sites $N$, i.e.~we have $\varphi_x$ for $\frac{x}{\Delta x} \in \left\lbrace0, 1, \dots, N - 1\right\rbrace$. We employ a discrete cosine transform of the individual realizations to obtain profiles $\tilde{\varphi}_{p_\text{lat}}$ and later translate the results to the Fourier transform. The lattice momentum takes the values ${p_\text{lat}} = j \frac{2\pi}{\Delta x N}$ with $j \in \left\lbrace -\frac{N}{2}, \dots, \frac{N}{2}-1 \right\rbrace$. We correct for some artifacts of the discrete transform at large momenta by considering physical momenta
\begin{align}
p_\text{phys} = \frac{2}{\Delta x} \sin \left(\frac{p_\text{lat} \Delta x}{2}\right) \; .
\end{align}
The two- and four-vertex densities (normalized to have units of $1/L$ with $L=N\Delta x$) from the main text are obtained by
\begin{subequations}
\begin{align}
\Gamma^{(2)}_p &= \frac{L}{\left\langle \left| \varphi_{p_\text{phys}} \right|^2 \right\rangle_{\text{c}}} \;, \\
\Gamma^{(4)}_p &= - \frac{L^3\left\langle \left| \varphi_{p_\text{phys}} \right|^4 \right\rangle_{\text{c}}}{\left\langle \left| \varphi_{p_\text{phys}} \right|^2 \right\rangle_{\text{c}}^4} \; .
\end{align}
\end{subequations}
Here, the expectation values are
\begin{subequations}
\begin{align}
	\left\langle \left| \varphi_{p} \right|^2 \right\rangle_{\text{c}} &= \left\langle \varphi_p \varphi_{-p} \right\rangle_{\text{c}} = 2\left\langle \tilde{\varphi}^2_p  \right\rangle_{\text{c}} \;, \\
		\left\langle \left| \varphi_{p} \right|^4 \right\rangle_{\text{c}} &= \left\langle \varphi_p \varphi_p \varphi_{-p} \varphi_{-p} \right\rangle_{\text{c}} = \frac{8}{3}\left\langle \tilde{\varphi}^2_p \right\rangle_{\text{c}} \;,
\end{align}
\end{subequations}
where the index $\text{c}$ indicates connected correlators according to Eq.~\eqref{eq:howto_connected}.

\subsection{One-loop corrections}
\label{app:One-loop}

To obtain the one-loop correction to the effective action, we approximate
\begin{widetext}
\begin{subequations}
	\begin{align}
	\beta K[\varphi, \Phi] &= \beta \left(H[\varphi+\Phi] - H[\Phi] - \frac{\delta H[\Phi]}{\delta \Phi_x} \varphi_x \right)\\
	&= \frac{\lambda_T}{4} \int_x \left\lbrace \frac{1}{2} \left(\partial_x \varphi_x\right)^2 - \frac{1}{\ell_J^2} \cos \Phi_x \left[-1 + \cos \varphi_x\right] - \frac{1}{\ell_J^2} \sin \Phi_x \left[\varphi_x - \sin  \varphi_x\right] \right\rbrace
	=  \frac{1}{2} \varphi_\mathbf{x} G^{-1}_{\mathbf{x},\mathbf{y}}[\Phi] \varphi_\mathbf{y} + \mathcal{O}\left(\varphi^3\right)
	\end{align}
\end{subequations}
\end{widetext}
with $G^{-1}_{xy} [\Phi] = \frac{\lambda_T}{4} \left[-\partial_x^2 + \frac{1}{\ell_J^2} \cos \left( \Phi_x\right)\right] \delta(x-y) $. From Eq.~\eqref{eq:explicit_one-loop}, we then have
\begin{align}\label{eq:one_loop}
	\Gamma'^{\text{,one-loop}}_\beta[\Phi] = -\frac{1}{2} \sum_{n=1}^{\infty} \frac{1}{n}\left(\frac{\lambda_T}{4} \frac{2}{\ell_J^2}\right)^n \text{Tr}  \left\lbrace G_0  \sin^2\left(\frac{\Phi}{2}\right)\right\rbrace^n \;,
\end{align}
where 
\begin{align}
G_{0,xy} = \int \frac{dp}{2 \pi} \, e^{ip(x-y)} G_{0,p}
\end{align} 
with \mbox{$G_{0,p} = \frac{4}{\lambda_T}/(p^2 + 1/\ell_J^2)$}. 

The one-loop corrections to the 1PI vertices are now obtained by taking derivatives of Eq.~\eqref{eq:one_loop}, evaluated at $\Phi=0$ in the symmetric case. Explicitly, we find at second order
\begin{align}
	\Delta \Gamma^{(2),\text{one-loop}}_{xy} &= -\frac{1}{2\ell_J^2}  G_{0,xx} \delta(x-y)
\end{align}
and at fourth order
\begin{align}
\Delta \Gamma^{(4),\text{one-loop}}_{xyzw} &= \frac{1}{2\ell_J^2} G_{0,xx} \delta(x-y) \delta(x-z) \delta(x-w) \nonumber\\
&- \frac{1}{2\ell_J^4} \left[G_{0,xy} G_{0,zw} \delta(x-z) \delta(y-w) \right. \nonumber \\
& \qquad\qquad \left.+ \text{2 perm.}\right]\; .
\end{align}
The involved loop integrals are given by
\begin{subequations}
\begin{align}
 \int_{-\infty}^{\infty} \frac{dq}{2\pi} \frac{1}{q^2 + 1/\ell_J^2} &= \frac{\ell_J}{2} \;,\\
 \int_{-\infty}^{\infty} \frac{dq}{2\pi} \frac{1}{q^2 + 1/\ell_J^2}\frac{1}{(p-q)^2 + 1/\ell_J^2} &= \frac{\ell_J}{(2/\ell_J)^2 + p^2}\;,
\end{align}
\end{subequations}
which results in the expressions given in the main text
\begin{subequations}
\begin{align}
	\Gamma^{(2),\text{one-loop}}_{p} &=  \frac{\lambda_T}{4}\left(p^2 + \frac{1}{\ell_J^2}\right)-\frac{1}{4\ell_J} \; , \\
	\Gamma^{(4),\text{one-loop}}_p &= - \frac{\lambda_T}{4 \ell_J^2} - \frac{1}{8 \ell_J^3} \frac{1}{p^2 + 1/\ell_J^2} \; .
\end{align}
\end{subequations}
Note that the one-loop correction is indeed of order $\mathcal{O}\left(\epsilon_{th}\right) = \mathcal{O}\left(1/Q\right) = \mathcal{O}\left(\ell_J /\lambda_T\right) $ compared to the tree-level approximation.


\bibliography{references}

\end{document}